\documentclass[aps,prd,superscriptaddress,notitlepage,11pt,final,longbibliography]{revtex4-1}

\usepackage{amsmath,amssymb,mathrsfs}
\usepackage{subfigure}
\usepackage{epsf}
\usepackage{epsfig}
\usepackage[usenames,dvipsnames]{xcolor}
\usepackage{bbm}
\usepackage{color}
\usepackage{comment}
\usepackage{cleveref}
\usepackage{comment}
\usepackage{natbib}

\graphicspath{{./fig/}}

\newcommand{\be}{\begin{equation}}
\newcommand{\ee}{\end{equation}}
\newcommand{\bea}{\begin{eqnarray}}
\newcommand{\eea}{\end{eqnarray}}

\newcommand{\eqn}{\begin{eqnarray}}
\newcommand{\eqnx}{\end{eqnarray}}

\newcommand{\pp}{\partial}

\newcommand{\rr}{\right}
\renewcommand{\ll}{\left }

\newcommand{\udagu}{1 + u^{\dagger} \cdot u}

\newcommand{\udag}{u^{\dagger}}

\usepackage{physics}
\numberwithin{equation}{section}

\begin{document}

\title{Gauged compact $Q$-balls and $Q$-shells in a multicomponent $CP^N$ model}


\author{P. Klimas~}
\email{pawel.klimas@ufsc.br}
\affiliation{Departamento de F\'isica, Universidade Federal de Santa Catarina, Campus Trindade, 88040-900, Florian\'opolis-SC, Brazil}


\author{L.C. Kubaski~}
\email{luizckubaski@gmail.com}
\affiliation{Departamento de F\'isica, Universidade Federal de Santa Catarina, Campus Trindade, 88040-900, Florian\'opolis-SC, Brazil}

\author{N. Sawado~}
\email{sawadoph@rs.tus.ac.jp}
\affiliation{Department of Physics and Astronomy, Tokyo University of Science, Noda, Chiba 278-8510, Japan}

\author{S. Yanai}
\email{yanai@toyota-ct.ac.jp}
\affiliation{Department of Natural Sciences, National Institute of Technology, Toyota College, Toyota, Aichi 471-8525, Japan}

\begin{abstract}
We study a multicomponent $CP^N$ model's scalar electrodynamics. The model contains $Q$-balls and $Q$-shells, 
which are nontopological compact solitons with time dependency $e^{i\omega t}$. 
Two coupled $CP^N$ models can decouple locally if one of their $CP^N$ fields takes the vacuum value.  
Because of the compacton nature of solutions, $Q$-shells can shelter another compact $Q$-ball or $Q$-shell within 
their hollow region.  Even if compactons do not overlap, they can interact through the electromagnetic field. 
We investigate how the size of multicompacton formations is affected by electric charge, 
with a focus on structures with nonzero or zero total net charge. 

\end{abstract}

\maketitle

\section{Introduction}
\label{sec:intro}
$Q$-balls and $Q$-shells are stationary field structures with finite energy in scalar field models. 
They are nontopological solitons with a distinct temporal dependency given by the factor $e^{i\omega t}$, \cite{Lee:1991ax, Coleman:1985ki}, 
which enables one to circumvent the Derrick argument and obtain stable field configurations. Although the scalar field is time dependent, the stationary nature of solutions is indicated by the time independence of the energy density of $Q$-ball and $Q$-shell configurations.

In contrast to topological scalar theories, $Q$-balls and $Q$-shells can be found in models with nondegenerate minima of the self-interaction potential. Traditional scalar field models contain quadratic potentials near their minima. In such a  case, the $Q$-ball energy density 
must cover a three-dimensional space quickly enough to keep the total energy finite. The presence of conserved charges is required due to the 
nontopological nature of $Q$-solitons. Such charges are known as Noether charges because they are associated with global continuous symmetries on a target space. The finiteness of energy and the conservation of Noether charges are insufficient for $Q$-ball stability. Indeed, these values should obey the relation $E\sim |Q|^{\alpha}$, where $\alpha<1$ guarantees that spontaneous splitting of a single $Q$-ball into two smaller $Q$-balls 
with charges $Q_1$ and $Q_2$ is not energetically beneficial.

$Q$-balls and $Q$-shells have attracted a lot of interest as spherical effective models of material media. 
One possible application for such systems is a boson star~\cite{Friedberg:1986tq, Jetzer:1991jr, Liebling:2012fv, Kleihaus:2009kr, Kleihaus:2010ep},
which necessitates the coupling of a scalar field with gravity, as well as possibly with an electromagnetic field. The presence of a special class of scalar field theories that support the compact nature of $Q$-balls and $Q$-shells is very important in this context. 
Because of their compactness, the energy density of the scalar field is strictly zero outside of a finite-sized support. 

The complex scalar field model permits compact $Q$-ball solutions when the field potential features a sharp minimum. This was specifically demonstrated in Ref.~\cite{Arodz:2008jk} using a V-shaped potential given by $V\sim |\phi|$. Historically, the concept of compactons (solitary waves with compact support) was first established in models exhibiting nonlinear dispersion \cite{Rosenau:1993zz, PhysRevLett.73.1737}, predating their discovery in models with nonanalytical potentials such as the V-shaped potential. Although the V-shaped potential may seem like an artificial construction at first glance, its physical relevance was later established by Adam \textit{et al.} \cite{Adam:2017pdh, Adam:2017srx} in the context of the Skyrme model and its BPS submodels, where it was shown to support compact oscillons. The analysis of solutions within the Skyrme model and its low-dimensional variations remains highly significant as a theoretical approach for modeling nuclear matter \cite{Adam:2010fg, Adam:2010ds, Gisiger:1996vb}.

The model utilizing this V-shaped potential has also been explored in the context of its coupling to the electromagnetic field \cite{Arodz:2008nm}. Furthermore, this model supports $Q$-shell solutions, which are distinctive spherical scalar field configurations featuring an empty interior cavity; this hollow structure arises as a consequence of the repulsive interaction between the field's electric charges.

Models that do not contain an electromagnetic field but require the presence of numerous scalar field components support the existence of $Q$-shell solutions. This type of solution has been demonstrated in \cite{Klimas:2017eft} for a $CP^N$ model with sharp potential. The compact gravitational charged $Q$-balls and $Q$-shells were reported in \cite{Sawado:2020ncc}. We recently found \cite{Klimas:2021eue} that more than one $CP^N$ model can be considered.  Compact excitations in each $CP^N$ component of the two coupled $CP^N$ models might differ significantly in their radial size.
 When the excitation of one $CP^N$ component overlaps with the vacuum solution of the second $CP^N$ component, the models decouple locally. Such solutions are especially intriguing for $Q$-shells since they allow for the existence of a $Q$-ball and $Q$-shell inside the external $Q$-shell's empty zone. In the absence of other fields, nonoverlapping compactons do not interact at all. However, when electrically charged matter fields are considered, the interaction is manifested by electromagnetic fields that extend beyond the compact spherical objects. The present work focuses on this issue. This would be the first stage before incorporating gravity further. We will look at how the presence of an electrical charge affects the radial size of multi $Q$-ball and $Q$-shell configurations.

The paper is structured as follows. In Sec. \ref{sec:sec1}, we show the multicomponent $CP^N$ model and its parametrization. 
In Sec. \ref{sec:sec2}, we explore the gauged form of the model, energy density, 
Noether charges, and electric charges of field configurations. 
Section \ref{sec:sec3} is devoted to numerical solutions, and Sec. \ref{sec:sec4} is devoted to a discussion and summary.

\section{The $CP^N$ model  and some other related models}
\label{sec:sec1}
The most common parametrization of the $CP^N$ model is based on the $N+1$ component complex vector ${\cal Z}=({\cal Z}_1,{\cal Z}_2,\ldots, {\cal Z}_{N+1})$, which satisfies the constraint ${\cal Z}^{\dagger}\cdot {\cal Z}=1$. The Lagrangian of the model is as follows:
\be
{\cal L}=\lambda_0\; {\cal D}^{\mu}{\cal Z}^{\dagger}\cdot {\cal D}_{\mu}{\cal Z}\label{lagr1}
\ee
where $\lambda_0$ is  a dimensional constant and ${\cal D}_{\mu}{\cal Z}\equiv \partial_{\mu }{\cal Z}-({\cal Z}^{\dagger}\cdot\partial_{\mu}{\cal Z}){\cal Z}$. To incorporate the constraint, new fields may be defined as $u_k\equiv\frac{{\cal Z}_k}{{\cal Z}_{N+1}}$, $k=1,2,\ldots,N$, 
which results in
\[
{\cal Z}=\frac{{\cal Z}}{\sqrt{{\cal Z}^{\dagger}\cdot {\cal Z}}}=\frac{{\cal Z}_{N+1}}{|{\cal Z}_{N+1}|}\frac{(u_1,u_2,\ldots,1)}{\sqrt{1+|u_1|^2+|u_2|^2+\ldots+|u_N|^2}}
\]
where the phase factor $\frac{{\cal Z}_{N+1}}{|{\cal Z}_{N+1}|}=e^{i\phi_{N+1}}$ is set to be unity. 
The Lagrangian \eqref{lagr1}, parametrized by $u_k$ fields, has the following form
\be
{\cal L}=\lambda_0\; \frac{(1+u^{\dagger}\cdot u)\partial^{\mu}u^{\dagger}\cdot\partial_{\mu}u-(\partial^{\mu}u^{\dagger}\cdot u)(u^{\dagger}\cdot\partial_{\mu}u)}{(1+u^{\dagger}\cdot u)^2}.\label{lagr2}
\ee
In this parametrization, the $CP^N$ field equations are as follows
\[
\partial^{\mu}\partial_{\mu}u_k-\frac{2}{1+u^{\dagger}\cdot u}(u^{\dagger}\cdot\partial^{\mu}u)\partial_{\mu}u_k=0.
\]
This equation can be solved with the help of rational maps~\cite{Zakrzewski1989LowdimensionalSM}. As shown in~\cite{Ferreira:2011nd, Ferreira:2011jy} the $CP^N$ model has some vortex solutions in 3+1 dimensions. Such vortex solutions have also been obtained in the extended Skyrme-Faddeev (SF) model, which, in addition to the Skyrme term, incorporates several quartic terms~\cite{Ferreira:2008nn, Ferreira:2010jb}. In reality, the extended SF model is equivalent to the $CP^N$ model with quartic terms included. When the target space of the model is $CP^1\sim SU(2)/U(1)$, the model is often parametrized in terms of a three-component unit vector, $\vec n\cdot\vec n=1$. In this scenario, the field $u_1\equiv u$ may be associated with $\vec n$ via stereographic projection 
\[
\vec n=\left(\frac{u+u^*}{1+|u|^2},-i\frac{u-u^*}{1+|u|^2},\frac{|u^2|-1}{1+|u|^2}\right).
\]

For a larger number of fields, i.e. when the target space is of the type $CP^N\sim SU(N+1)/SU(N)\otimes U(1)$, 
the principal variable $X$ provides a convenient parametrization~\cite{ Ferreira:1998zx}. This parametrization was used in the case of the extended SF model, discussed in  \cite{Ferreira:2011jy}, as well as in some other related publications. Many technical aspects of this parametrization can be found in \cite{ Ferreira:1998zx}, including the parametrization of group elements $g\in SU(N+1)$ in terms of fields $u_k$, where $k=1,2,\ldots,N$. Here, we only recall the result 
\be
X(g)=g^2 = \begin{pmatrix}
    \mathbb{I}_{N\times N} & 0 \\
    0 & -1
        \end{pmatrix} + \frac{2}{\udagu}
        \begin{pmatrix}
            -u \otimes \udag & i u\\
            i \udag & 1
        \end{pmatrix}\label{Xparam}
\ee
where $\mathbb I_{N\times N}$ is the identity $N\times N$ matrix. The $CP^N$ Lagrangian, which is one of the terms in the extended SF model, has the following form in terms of the principal variable $X$
\be
{\cal L} = - \frac{M^2}{2}\Tr(X^{-1}\pp_{\mu}X)^2\label{lagrcpnX}
\ee
where $M$ is a dimensional constant.
It is worth noting that, in terms of fields $u_k$ introduced in \eqref{Xparam}, the Lagrangian \eqref{lagrcpnX} can be identified with \eqref{lagr2} for $\lambda_0=4M^2$. There are three types of quartic terms that can be included in the model, namely the Skyrme term ${\rm Tr}([X^{-1}\partial_{\mu}X,X^{-1}\partial_{\nu}X])^2$ and  $\left[{\rm Tr}(X^{-1}\partial_{\mu}X)^2\right]^2$ and  $ \left[{\rm Tr}(X^{-1}\partial_{\mu}XX^{-1}\partial_{\nu}X)\right]^2$.

The inclusion of the potential leads to the possibility of an additional extension. This model modification method is particularly useful for a baby-Skyrme model, 
which provides solutions for a typical {\it old potential} \cite{Piette:1994ug, Piette:1994mh} and some of its generalizations \cite{Hen:2007in, Karliner:2009at}. Furthermore, baby skyrmions have compact supports for a particular choice of
potential \cite{Adam:2009px}. Some nontrivial solutions for the SF model in 3+1 dimensions have also been obtained. According to \cite{Ferreira:2011mz}, the extended $CP^1$ SF model with an appropriately chosen potential has exact vortex solutions 
that are the product of two functions, $f(x^1\pm ix^2)$ and $g(x^0\pm x^3)$. Further research on vortex solutions in the extended SF model is presented in \cite{Klimas:2012aw, Amari:2015sva}. It should be noted that such vortices are likewise $CP^N$ model solutions without potential \cite{Ferreira:2011nd}. In other words, identical field configurations are solutions to different models.

The vortex solution is definitely intriguing, but it has one drawback. Its total energy is infinite due to the infinite vortex length. 
Another type of dimensional reduction can be used to obtain solutions with finite total energy in 3+1 dimensions. 
Using Ferreira's ansatz, the $CP^N$ term (and its quartic extensions in the case of the SF model) can be reduced to a single 
radial equation. When $N$ is odd, ($N=2l+1$), each of the $2l+1$ complex fields can be parametrized as $u_m \sim f(r)Y_{lm}(\theta,\phi)$, 
where $r,\theta,\phi$ are spherical coordinates and the fields are indexed by $m=-l,\ldots, l$ instead of $k=1,2,\ldots,2l+1$.  
Unfortunately, the resulting ordinary equation for $f(r)$, which is of the form
\be
f''+\frac{2}{r}f'-\frac{l(l+1)}{r^2}f-\frac{2ff'^2}{1+f^2}=0\label{cpneqnopot}
\ee
only has a trivial solution $f(r)=0$ (or a constant for $l=0$). We shall demonstrate it for $l=0$. 
When $f(r)$ is expanded around $r=0$, i.e. $f(r)=a_0+a_1r+a_2 r^2+\ldots$, and substituted into Eq.~\eqref{cpneqnopot}, it yields 
\[
\frac{2 a_1}{r}+\left(6 a_2-\frac{2 a_0 a_1^2}{a_0^2+1}\right)+\left(\frac{2
   \left(a_0^2-1\right) a_1^3-8 \left(a_0^3+a_0\right) a_1
   a_2}{\left(a_0^2+1\right){}^2}+12 a_3\right) r+O\left(r^2\right)=0.
\]
One solution would be if all coefficients multiplying powers of $r$ were equal to zero. However, it is necessary that the free coefficients satisfy specific relations. Looking at the expansion expression, we can see that the singular term can be eliminated by selecting $a_1=0$. Other odd terms can also be removed by setting higher odd coefficients $a_k=0$, which results in an expression with only even terms:
 \[
 6 a_2+\left(20 a_4-\frac{8 a_0 a_2^2}{a_0^2+1}\right) r^2+O\left(r^3\right)=0.
 \]
 The resulting formula shows that the choice $a_2=0$ implies $a_4=0$ and so on. Ultimately, we are left with a trivial constant solution $f(r)=0$. To avoid this problem, a suitable potential might be included in the model. The potential must allow for the cancellation of the constant term $6a_2$ in the field equation. Because the potential contributes to the field equation via its derivative with respect to the field, this derivative must be constant around the vacuum solution $f=0$. It may appear unusual, particularly when compared to the great majority of field potentials with vanishing derivatives at potential minima. Fortunately, it turns out that there is a family of field theories with sharp (nonanalytic) potentials at the minima \cite{Arodz:2005gz}. Such potentials are known as V-shaped potentials. The signum-Gordon model, \cite{Arodz:2007ek}, is the simplest model with V-shaped potential $V(\phi)=|\phi|$.  Despite its apparent simplicity, the model has extremely complex dynamics. The nonanalytic potentials are effectively included in the $CP^N$ model, where they enable compact $Q$-balls and $Q$-shells with finite total energy, according to \cite{Klimas:2017eft}. Because this subject is strongly related to our current work idea, we will comment on certain specifics below. The following Lagrangian describes the model presented in \cite{Klimas:2017eft}:
 \be
 {\cal L} = - \frac{M^2}{2}\Tr(X^{-1}\pp_{\mu}X)^2-\mu^2 V(X)\label{singleCPN}
 \ee
 where the potential reads
 \be
 V(X) = \frac{1}{2}\ll[ \Tr(\mathbb{I}-X) \rr]^{\frac{1}{2}} = \ll( \frac{\udag \cdot u}{\udagu}  \rr)^{\frac{1}{2}}.
 \ee
To find solutions avoid Derrick's theorem, one needs to modify Ferreira's ansatz to include the temporal dependency of solutions provided by the standard $Q$-ball factor $e^{i\omega t}$. Indeed, as noted in \cite{Klimas:2017eft}, there are no solutions for a static sector $\omega=0$. The solutions are only available for $\omega>\omega_\textit{min}$. The $Q$-ball and $Q$-shell solutions have the form
 \be
 u_m(t,r,\theta,\phi) = \sqrt{\frac{4 \pi}{2l+1}}f(r) Y_{lm}(\theta, \phi)e^{i \omega t},\label{ansatz1}
 \ee
 where the normalization factor is such that $u^{\dagger}\cdot u=f(r)^2$.
The solutions are compact, which means they are only nontrivial (different from the vacuum solution) on specific compact supports. Outside of this support, the fields take their vacuum value $u_k=0$. The extended $CP^N$ model is distinguished by the presence of shell solutions in the absence of an electromagnetic field. It should be emphasized that in traditional $Q$-ball solutions, the shell form emerges as a result of a repulsive force caused by electric charges.

The importance of such solutions stems from their applicability to boson stars, as commented in \cite{Friedberg:1986tq, Jetzer:1991jr, Liebling:2012fv, Kleihaus:2009kr, Kleihaus:2010ep}. The generalization of the model is investigated in \cite{Sawado:2020ncc, Sawado:2021rsc} where the authors analyze the features of compact solutions that are connected to gravity and an electromagnetic field. Special consideration should be given to the solution that includes a spherical massive body in the center of the spherical shell. Such a body could be a nonspinning black hole, in particular. This is referred to as a {\it harbor solution}.

We recently looked into another interesting case in which the empty space inside a $Q$-shell contains another $Q$-ball or $Q$-shell compacton.  
This type of solution exists in field theoretical models composed of two or more $CP^N$ models.  As a result, we refer to these models as {\it multicomponent models} \cite{Klimas:2021eue}. The Lagrangian \eqref{singleCPN} defines each $CP^N_a$ component, $a=1,2,\ldots,n$. Each such component is labeled as follows: 
\[
{\cal L}_a={\cal L}_a(X_a,\partial_{\mu}X_a;N_a,M_a,\mu_a), \qquad a=1,2,\ldots,n
\]
in which $M_a$ and $\mu_a$ are coupling constants. As a result, the Lagrangian 
\be
{\cal L}_{CP^N}=\sum_{a=1}^n{\cal L}_a+{\cal L}_{int}
\ee
is the multicomponent model, where ${\cal L}_{int}$ is the interaction term comprising the multicomponent fields $a\neq b$. The model would decouple into numerous distinct 
$CP^N$ models with V-shaped potentials if this term was not present. One of the simpler options for the interaction term is to take the products of the powers of the nonanalytical potentials $V(X_a)$, i.e.,
\be
{\cal L}_{int}=-\lambda\sum_{a\neq b}W(X_a,X_b)\label{intlagr}
\ee where
\begin{align}
W(X_a,X_b)&=V(X_a)^{2 \alpha}V(X_b)^{2 \beta}= \left( \frac{1}{4}\Tr({\mathbb {I}} - X_a) \right)^{\alpha} \left( \frac{1}{4}\Tr({\mathbb {I}} - X_b) \right)^{\beta}
\label{Wint}
\end{align}
with $\alpha, \beta \ge 1$.  The model decouples locally in regions where compactons do not overlap as a result of this choice. In the two-component model, for example, if the field of the first component takes a vacuum value $u^{(1)}_m=0$, i.e. $X_1={\mathbb I}$, resulting in a zero interaction term, the field of the second component is locally determinable by the Lagrangian ${\cal L}_2$. The same holds true for the first component when the field of the second component is set to zero. The harbor solution can be included in this model. More generic compact solutions including interacting fields from multiple model components have been developed.
In the parametrization provided by the $u_m^{(a)}$ fields, the Lagrangian of a multicomponent model takes the following form: 
\begin{align}
{\cal L}_{CP^N}=\sum_{a=1}^n {\cal L}^{(a)}-\sum_{a\neq b}\lambda\;W(u^{(a)},u^{(b)}),\label{multiLu}
\end{align}
where
\begin{align}
&{\cal L}^{(a)}={\cal L}_{kin}^{(a)}-\mu^2_a V(u^{(a)})\label{La}\\
&{\cal L}_{kin}^{(a)}=4M_a^2\;\partial_{\mu}u^{\dagger(a) }\cdot\widetilde \Delta^{2(a)}\cdot\partial^{\mu}u^{(a)}\label{Lkin}\\
&\widetilde \Delta^{2(a)}_{ij}\equiv\frac{(1+u^{\dagger (a)}\cdot u^{(a)})\delta_{ij}-u_i^{(a)}u_j^{*(a)}}{(1+u^{\dagger (a)}\cdot u^{(a)})^2},\label{Delta2}\\
&V(u^{(a)})=\left(\frac{u^{\dagger (a)}\cdot u^{(a)}}{1+u^{\dagger (a)}\cdot u^{(a)}} \right)^{\frac{1}{2}},\label{potV}\\
&W(u^{(a)},u^{(b)})=V(u^{(a)})^{2\alpha}V(u^{(b)})^{2\beta}.\label{potW}
\end{align}

In order to obtain solutions for boson stars, this model should be coupled with gravity and, if necessary, also with electromagnetism. In this research, we investigate the case of $Q$-spheres and $Q$-shells with an electric charge but without considering the gravitational interaction. The model is defined in the next section, followed by the solution of the field equations and the analysis of compactons.

\section{Gauged $CP^N$ multicomponent model}
\label{sec:sec2}
\subsection{Action and field equations}
We consider the Lagrangian of the multicomponent model \eqref{multiLu} minimally coupled to the electromagnetic field. We assume that the Lagrangian is invariant with respect to the local transformation
\be
u^{(a)}_j\rightarrow e^{iq^{(a)}\Lambda(x)}u^{(a)}_j,\qquad j=1,2,\ldots, N_a
\ee
where $\Lambda(x)$ is an arbitrary function of spacetime coordinates and $q^{(a)}$ are constant parameters.
This is possible provided that partial derivatives $\partial_{\mu}u^{(a)}_j$ in \eqref{Lkin} are replaced by covariant derivatives
\be
D_{\mu} u^{(a)}_j = \partial_{\mu} u^{(a)}_j - i e q^{(a)} A_{\mu} u^{(a)}_j,
\ee
where  $A_{\mu}$ is the connection transforming according to
\be
A_{\mu}(x)\rightarrow A_{\mu}(x)+\frac{1}{e}\partial_{\mu}\Lambda(x).\label{gaugesymmetry}
\ee
The constant $e$ stands for the elementary value of the electric charge.
The covariant derivative is transformed in the same way as  $u^{(a)}_j$ fields i.e.
\be
D_{\mu} u^{(a)}_j\rightarrow e^{iq^{(a)}\Lambda(x)}D_{\mu} u^{(a)}_j.
\ee
The connection field can be interpreted as a physical field. Its contribution to the Lagrangian is given by
\be
{\cal L}_{EM}=-\frac{1}{4}F_{\mu\nu}F^{\mu\nu}, \qquad F_{\mu\nu}\equiv\partial_{\mu}A_{\nu}-\partial_{\nu}A_{\mu}.\label{emlagr}
\ee
In this paper we consider the model given by the following action
\be
S=\int d^4x\sqrt{-g}\big({\cal L}_{CP^N}+{\cal L}_{EM}\big),\label{action}
\ee
where \eqref{multiLu}--\eqref{potW} give the Lagrangian ${\cal L}_{CP^N}$ with scalar fields substituted with covariant ones,
\begin{align}
{\cal L}_{kin}^{(a)}=4M_a^2\;\Big(D_{\mu}u^{\dagger(a) }\cdot\widetilde \Delta^{2(a)}\cdot D^{\mu}u^{(a)}\Big),\label{LkinD}
\end{align}
 and \eqref{emlagr} gives the ${\cal L}_{EM}$.

To obtain the scalar field equations, we vary the action with respect to $u^{\dagger(a)*} $. Variation with respect to $u^{\dagger(a) }$ fields yields equations that are complex conjugates of equations produced by variation with respect to $u^{\dagger(a)*} $. In turn, the variation 
regarding the four-potential $A_{\mu}$ leads to Maxwell's equations. The field equations are as follows
\begin{align}
    \label{em:el_equation4}
    &\frac{1}{\sqrt{-g}}D_{\mu} \left( \sqrt{-g} D^{\mu} u^{(a)}_j \right) - 2\frac{\left(u^{(a) \dagger} \cdot D_{\mu} u^{(a)}\right) D^{\mu} u^{(a)}_j}{1+u^{(a)\dagger}\cdot u^{(a)}} +\nonumber \\
   & \hspace{0.5cm}+ \frac{1}{4}\Big(1+u^{(a)\dagger}\cdot u^{(a)}\Big) \sum_{l=1}^{N_a} \left\{ \left(\delta_{jl} + u^{(a)}_j u^{(a)*}_l\right)\left[\frac{\mu_a^2}{M_a^2} \frac{\delta V_a}{\delta u^{(a)*}_l} + \frac{\lambda}{M_a^2}\frac{\delta W}{\delta u^{(a)*}_l}\right]\right\} = 0,
   \\
   &\frac{1}{\sqrt{-g}}\partial_{\mu} \left( \sqrt{-g} F^{\mu \nu} \right) 
    = - \sum_{a=1}^n \frac{4 i e q^{(a)} M_a^2}{(1+u^{(a)\dagger}\cdot u^{(a)})^2} \left[ u^{(a) \dagger} \cdot D^{\nu} u^{(a)} - D^{\nu}u^{(a)\dagger} \cdot u^{(a)}  \right]\label{em:max_equations2_q}.
\end{align}

\subsection{The $Q$-ball ansatz}
Though the technical difficulties are significantly greater, the model can be freely modified to include any number of components.  
We focus on a model with only two $CP^N$ components, $u_j\equiv u_j^{(1)}$ and $v_j\equiv u^{(2)}_j$, to reduce complexity. 
When seeking spherically symmetric solutions (based solely on radial functions) inside both components of the model, the same ansatz 
for both components \eqref{ansatz1} can be used. 
The number of scalar fields for each $CP^N$ model is denoted as $N_1=2l_1+1$ and $N_2=2l_2+1$. 
We also use a more convenient field numbering scheme, substituting $u_j$, $v_k$, 
$j,k=1,2,\ldots$ with $u_{m_1}$, $m_1=\{-l_1,-l_1+1,\ldots, l_1-1,l_1\}$, and similarly $v_{m_2}$, 
$m_2=\{-l_2,-l_2+1,\ldots,l_2-1,l_2\}$. This notation is more appropriate for the  ansatz
\begin{align}
&u_{m_1}(t,r,\theta,\phi) = \sqrt{\frac{4 \pi}{2 l_1 + 1}}f(r)Y_{l_1,m_1}(\theta,\phi)e^{i \omega_1 t}, 
    \label{em:eq_um1}
    \\
    &v_{m_2}(t,r,\theta,\phi) = \sqrt{\frac{4 \pi}{2 l_2 + 1}}g(r)Y_{l_2,m_2}(\theta,\phi)e^{i \omega_2 t}.
    \label{wm:eq_um2}
\end{align}
 In addition, we select a gauge in which the four-potential $A_{\mu}$ just has a temporal component that is a function of the radial coordinate
 \begin{align}
    &A_{\mu}(t,r,\theta,\phi) = (A_t (r),0,0,0).\label{ansatzA}
\end{align}
This means that the spatial components of the covariant derivative coincide with the spatial components of the partial derivative $D_i\equiv \partial_i$ .
When \eqref{em:eq_um1}, \eqref{wm:eq_um2}, and \eqref{ansatzA} are put into  \eqref{em:el_equation4} and \eqref{em:max_equations2_q},
the following form of the equations are obtained:
\begin{align}
    &\Sigma^{(1)}(r)=\frac{\widetilde \mu_1^2}{8}\sqrt{1+f^2}\;{\textrm{sgn}}(f)+\lambda_1\frac{\alpha}{4}\left(\frac{f^2}{1+f^2}\right)^{\alpha-1}\left(\frac{g^2}{1+g^2}\right)^{\beta}f,\label{em:eq1f}\\
    &\Sigma^{(2)}(r)=\frac{\widetilde \mu_2^2}{8}\sqrt{1+g^2}\;{\textrm{sgn}}(g)+\lambda_2\frac{\beta}{4}\left(\frac{f^2}{1+f^2}\right)^{\alpha}\left(\frac{g^2}{1+g^2}\right)^{\beta-1}g\label{em:eq1g}\\
    &A_t'' + \frac{2}{r}A_t' + \rho(r)=0,\label{em:eq1a}
\end{align}
where
\begin{align}
\rho(r)&\equiv  8\left( M_1^2 e q^{(1)} b_1(r) \frac{f^2}{(1+f^2)^2} + M_2^2 e q^{(2)} b_2(r) \frac{g^2}{(1+g^2)^2}\right)\label{rhodens}\\
b_a(r)&\equiv\omega_a-eq^{(a)}A_t(r),\qquad a=1,2\label{ba}\\
\lambda_a&\equiv\frac{\lambda}{M_a^2},\qquad
\widetilde\mu_a\equiv \frac{\mu_a^2}{M_a^2}\nonumber
\end{align}
and where we have denoted by 
\begin{align}
    &\Sigma^{(1)}(r)\equiv f''+\frac{2}{r}f'+b_1^2(r)\frac{1-f^2}{1+f^2}f-2\frac{f f'^2}{1+f^2}-\frac{l_1(l_1+1)}{r^2}f,\label{em:Sigma-f}\\
    &\Sigma^{(2)}(r)\equiv g''+\frac{2}{r}g'+b_2^2(r)\frac{1-g^2}{1+g^2}g-2\frac{g g'^2}{1+g^2}-\frac{l_2(l_2+1)}{r^2}g\label{em:Sigma-g}
\end{align}
expressions that are derived from the component of the Lagrangian that does not contain potentials.
We omit the radial argument of $f(r)$ and $g(r)$ for simplicity.
The $b_a(r)$ are not independent functions because they are defined in terms of a single function $A_t(r)$, 
Additionally, in the limit $q^{(a)} \rightarrow 0$, we have $b_a \rightarrow \omega_a$ and we recover the equations corresponding to 
the multicomponent $CP^N$ model \cite{Klimas:2021eue}. Finally, we see that Eq. \eqref{em:eq1a} is merely a Gauss law $\nabla \cdot \vec E=\rho(r)$ with  electric charge density that is dependent on the gauge field. Indeed, by introducing $\vec E=-\nabla A_t$, this equation may be written as $\nabla^2A_t+\rho(r)=0$. A radial component of the electric field in this case is $E_r=-A'_t$

\subsection{Energy of a field configuration}
The energy of the field is composed of two components: the energy of matter described by the scalar field and the energy of the electromagnetic field. We employ the method of variation with respect to the metric tensor to determine the energy density, which ensures that the energy-momentum tensor $T^{\mu\nu}$ is symmetric. The component $T^{00}$ provides the energy density of the field configuration. We examine $T^{\mu\nu}$ as defined below
\[
\delta S=\frac{1}{2}\int d^4x\sqrt{-g}\;T_{\mu\nu}\delta g^{\mu\nu}
\]
where  \eqref{action} denotes the action $S$. Because our Lagrangian is not dependent on metric tensor derivatives, $T_{\mu\nu}$ has the form
\[
T_{\mu\nu}=2\frac{\delta {\cal L}}{\delta g^{\mu\nu}}-g_{\mu\nu}{\cal L},\qquad{\rm where}\qquad {\cal L}={\cal L}_{CP^N}+{\cal L}_{EM}.
\]
The energy-momentum tensor is the result of the addition of two tensors
\begin{align}
T_{\mu\nu}^{CP^N}&=4\sum_{a=1}^nM^2_a\bigg[2\Big(D_{\mu}u^{\dagger(a) }\cdot\widetilde \Delta^{2(a)}\cdot D_{\nu}u^{(a)}\Big)-g_{\mu\nu}g^{\alpha\beta}\Big(D_{\alpha}u^{\dagger(a) }\cdot\widetilde \Delta^{2(a)}\cdot D_{\beta}u^{(a)}\Big)\bigg]\nonumber\\
&+g_{\mu\nu}\bigg[\sum_{a=1}^n \mu^2_a V(u^{(a)})+\sum_{a\neq b}\lambda\;W(u^{(a)},u^{(b)}) \bigg],\\
T_{\mu\nu}^{EM}&=-F_{\mu}^{\beta}F_{\nu\beta}+\frac{1}{4}g_{\mu\nu}F_{\alpha\beta}F^{\alpha\beta}.
\end{align}
The chosen gauge gives $D_{0}u^{(a)}=ib_a u^{(a)}$, $D_{i}u^{(a)}=\partial_iu^{(a)}$. In addition, $F_{r0}=\partial_rA_t\equiv A'_t$ is a unique nonvanishing component of the electromagnetic field tensor. The component $T_{00}=T^{CP^N}_{00}+T_{00}^{EM}$ for the case of a two-component model reads
\begin{align}\label{T00}
T_{00}&=4M_1^2\bigg[\frac{b_1^2f^2+f'^2}{(1+f^2)^2}+\frac{l_1(l_1+1)}{r^2}\frac{f^2}{1+f^2}\bigg]+\mu_1^2\frac{f}{\sqrt{1+f^2}}\nonumber\\
&+4M_2^2\bigg[\frac{b_2^2g^2+g'^2}{(1+g^2)^2}+\frac{l_2(l_2+1)}{r^2}\frac{g^2}{1+g^2}\bigg]+\mu_2^2\frac{g}{\sqrt{1+g^2}}\\
&+\lambda\bigg(\frac{f^2}{1+f^2}\bigg)^{\alpha}\bigg(\frac{g^2}{1+g^2}\bigg)^{\beta}+\frac{1}{2}(A'_t)^2,\nonumber
\end{align}
where $b_a\equiv b_a(r)$, $f\equiv f(r)$, $g\equiv g(r)$ and $A_t\equiv A_t(r)$. As a result, the energy of the field configuration is as follows
\be
E=4\pi\int_0^{\infty} dr r^2 T_{00}.
\ee

\subsection{Noether charges}
The action \eqref{action} is invariant with respect to continuous global symmetry  $U(1)^{N_1}\otimes U(1)^{N_2}$
\begin{align}
    &u_{s} \xrightarrow{} u_{s} e^{i \alpha_s}, \qquad s=1,\dots, N_1 \nonumber\\
	&v_{p} \xrightarrow{} v_{p} e^{i \beta_p}, \qquad p=1,\dots, N_2.\nonumber\\
	& A_{\mu} \xrightarrow{} A_{\mu}.\nonumber
\end{align}
We examine infinitesimal transformations generated by $\delta \alpha_k$ and $\delta \beta _i$
\begin{align}
&u_{s}\xrightarrow{}(1+i\delta \alpha_s)u_s=u_s+\delta u_s\\
&v_{p}\xrightarrow{}(1+i\delta \beta_p)v_p=v_p+\delta v_p
\end{align}
which gives
\be
\frac{\delta u_k}{\delta\alpha_s}=iu_k\delta_{ks},\qquad \frac{\delta u^*_k}{\delta\alpha_s}=-iu^*_k\delta_{ks},\qquad k,s=1,\dots, N_1 \nonumber
\ee
and similarly
\be
\frac{\delta v_i}{\delta\beta_p}=iv_i\delta_{ip},\qquad \frac{\delta v^*_i}{\delta\beta_p}=-iv^*_i\delta_{ip}, \qquad i,p=1,\dots, N_2.\nonumber
\ee
The variation of the Lagrangian associated with the infinitesimal transformations has the form
\be
\delta {\cal L} =\sum_{s=1}^{N_1}\delta\alpha _s\;\partial^{\mu}J^{s, (1)}_{\mu}+\sum_{p=1}^{N_2}\delta\beta _p\;\partial^{\mu}J^{p, (2)}_{\mu}
\ee
where
\begin{align}
&J^{s, (1)}_{\mu}=\frac{\delta {\cal L}}{\delta(\partial^{\mu}u^*_k)}\frac{\delta u^*_k}{\delta\alpha_s}+\frac{\delta {\cal L}}{\delta(\partial^{\mu}u_k)}\frac{\delta u_k}{\delta\alpha_s}\qquad k,s=1,\dots, N_1 \nonumber\\
&J^{p, (2)}_{\mu}=\frac{\delta {\cal L}}{\delta(\partial^{\mu}v^*_i)}\frac{\delta v^*_i}{\delta\beta_p}+\frac{\delta {\cal L}}{\delta(\partial^{\mu}v_i)}\frac{\delta v_i}{\delta\alpha_p}\qquad i,p=1,\dots, N_2. \nonumber
\end{align}
are the conserved Noether currents. These currents have the following form
\begin{align}
J^{s, (1)}_{\mu}=-4M_1^2\,i\,\sum_{k=1}^{N_1}\left[u_s^*\widetilde\Delta^{2 (1)}_{sk}D^{\mu}u_k-D^{\mu}u^*_k\widetilde\Delta^{2 (1)}_{ks}u_s\right],\qquad s=1,\dots, N_1\\
J^{p, (2)}_{\mu}=-4M_2^2\,i\,\sum_{i=1}^{N_2}\left[v_p^*\widetilde\Delta^{2 (2)}_{pi}D^{\mu}v_i-D^{\mu}v^*_i\widetilde\Delta^{2 (2)}_{ip}v_p\right],\qquad p=1,\dots, N_2.
\end{align}
In the next step, we express these formulas using the dimensional reduction given by ansatz \eqref{em:eq_um1} and ansatz \eqref{wm:eq_um2}. 
We also replace the index $s=1,\ldots, N_1$ by $m_1=-l_1,\ldots, l_1$ and $p=1,\ldots, N_2$ by $m_2=-l_2,\ldots, l_2$ where $N_1=2l_1+1$ and $N_2=2l_2+1$. Below we present expressions only for the first $CP^N$ component because the expressions for the second $CP^N$ component have an analogous form. We obtain
\begin{align}
    &\widetilde{J}^{m_1,(1)}_t=8 b_1 \frac{(l_1-m_1)!}{(l_1+m_1)!}\frac{f^2}{(1+f^2)^2}(P_{l_1}^{m_1}(\cos\theta))^2\,,
	\label{em:currentt}\\
	&\widetilde{J}^{m_1 (1)}_\varphi=8m_1\frac{(l_1-m_1)!}{(l_1+m_1)!}\frac{f^2}{1+f^2}(P_{l_1}^{m_1}(\cos\theta))^2.
	\label{em:currentp}
\end{align}
The current $\widetilde{J}^{m_1,(1)}_{\mu}$ is conserved. This can be proved by direct calculation
\begin{align}
    \frac{1}{\sqrt{-g}}\partial_{\mu}\left( \sqrt{-g} g^{\mu \nu} \widetilde{J}^{m_1,(1)}_{\nu} \right)= \partial_t \widetilde{J}^{m_1,(1)}_{t} - \frac{1}{r^2\sin^2\theta}\partial_\varphi \widetilde{J}^{m_1,(1)}_{\varphi}=0
\end{align}
where the $\nu=t$ current component does not depend on time and  the $\nu=\varphi$ component does not depend on the angle $\varphi$. The Noether currents associated with the second component of the model have an analogous form. One just needs to replace $f(r)$ by $g(r)$, $m_1$ by $m_2$ and $l_1$ by $l_2$.
The conserved Noether charges read
\begin{align}
    Q_t^{(m_1)}&=\int d^3 x\sqrt{-g}\widetilde{J}_t^{m_1,(1)}
	=\frac{32\pi}{2l_1+1}\int^\infty_0dr r^2 \frac{ b_1 f^2}{(1+f^2)^2}\,.
	\label{NoetherCharge}
\end{align}
Note that they have exactly the same form for each value of $m_1$.

\subsection{Electric charges}
The electric charges are determined by the asymptotic behavior of the gauge field  (electrostatic potential). This field is a solution of ~(\ref{em:eq1a}). In fact, this equation is the Poisson equation satisfied by the electrostatic potential. The difficulty, however, is that the charge density is not a given function but depends on the potential. After integration, this equation leads to the relation between the electric field and the electric charge contained in a spherical region of radius $r$. Below we show that electric charges and Noether charges~(\ref{NoetherCharge}) of such field configurations are related. 
The equation of motion (\ref{em:eq1a}) can be cast in the form
\begin{align}
-(r^2A_t(r)')'=r^2\rho(r)
\end{align}
where $\rho(r)$ is given by \eqref{rhodens}. After integration over the spherical region,  we obtain
\begin{align}
-A_t(r)'=\frac{1}{r^2}\int_0^rdr'\,r'^2\rho(r')
\end{align}
which is exactly the radial component of the electrostatic field $E_r(r)=-A_t(r)'$. Note that the expression $\rho(r)$ given by \eqref{rhodens} vanishes in the region where both $CP^N$ components take their vacuum values i.e $f=0$ and $g=0$. 

We can use the Gauss law to determine the electric charges of compactons., namely,  $-A_t'(r)={\bar Q}/(4\pi r^2)$ for sufficiently large $r$. Of course, the expressions $b_1(r')$ and $b_2(r')$, which according to \eqref{ba} depend on $A_t(r')$, must be explicitly known in order to integrate the charge density. The function $A_t(r)$, together with $f(r)$ and $g(r)$, is obtained from numeric integration of the field equations. 

For the harbor-type solution composed of the nonoverlapping $Q$-ball, $0\leq r\leq R_1$, and the 
$Q$-shell, $R_2^\textrm{(in)}\leq r\leq R_2^\textrm{(out)}$,  the electric charges can be defined in the same way. 
The electric charge of the ball may be deduced from the gauge field $A_t(r)'=-\bar{Q}_1/(4\pi r^2)$ outside of the ball and it reads
\begin{align}
\bar{Q}_1&:=4\pi\int_0^rdr'\,r'^2\rho(r')\nonumber\\&=32\pi M_1^2eq^{(1)}\int_0^{R_1} dr'\,r'^2\,b_1(r')\frac{f(r')^2}{(1+f(r')^2)^2}\nonumber\\&
\equiv (2l_1+1)M_1^2eq^{(1)}Q_1,
\label{Charge1}
\end{align}
where $Q_1$ is the Noether charge \eqref{NoetherCharge}.
In addition, the net electric charge $\bar{Q}_\textrm{net}$ is defined by the gauged field $A_t'=-\bar{Q}_\textrm{net}/(4\pi r^2)$ integrating over the region inside the ball $r<R$ where $R>R_2^{\textrm{(out)}}$ is the outer radius of the shell. Since, the matter field functions $f(r')$ and $g(r')$ do not vanish only inside the ball $r<R_1$ and inside the shell $R_2^{\textrm{(in)}}<r<R_2^{\textrm{(out)}}$, one gets the electric net charge
\begin{align}
\bar{Q}_\textrm{net}&:=32\pi\,M_1^2eq^{(1)}\int_0^{R_1} dr'\,r'^2\,b_1(r')\frac{f(r')^2}{(1+f(r')^2)^2}\nonumber\\
&+32\pi\,M_2^2eq^{(2)}\int_{R_2^{\textrm{(in)}}}^{R_2^{\textrm{(out)}}}dr'\,r'^2\,b_2(r')\frac{g(r')^2}{(1+g(r')^2)^2}
\nonumber \\
&\equiv \bar{Q}_1
+\bar{Q}_2.
\end{align}
Similarly, the electric charge of the shell $\bar{Q}_2$ is defined by
\begin{align}
\bar{Q}_2&:=32\pi\,M_2^2eq^{(2)}\int_{R_2^\textrm{(in)}}^{R_2^\textrm{(out)}} dr' r'^2\,b_2(r')\frac{g(r')^2}{(1+g(r')^2)^2}\nonumber\\&
\equiv (2l_2+1)M_2^2eq^{(2)}Q_2.
\label{Charge2}
\end{align} 

The sign of the Noether charge \eqref{NoetherCharge} and the sign of $q^{(a)}$ determine the sign of electric charge. 
Note that $e>0$.  Any two scalar field configurations that differ only in the sign of the electric charge density should have the same radial profile functions. In standard electrostatics of pointlike electric charges, two charges that differ solely in sign have potentials that also differ in sign. The same can be said for any continuous spherical configuration that differs only in the sign of a distribution $\rho(r)$. Going back to scalar electrodynamics,  Poisson equation \eqref{em:eq1a} shows that changing the sign of the charge density $\rho(r)\rightarrow-\rho(r)$ changes the sign of the scalar potential $A_t\rightarrow -A_t$. To determine the sign of the topological charge, and consequently the sign of the electric charge, only single sign functions $b_a(r)$ can be utilized. We deal with more than one electrically charged compacton in general, which may cause $A_t$ to change sign depending on the radial coordinate; therefore, the expressions $b_a(r)$ are not considered to be single sign functions. As a result, it is more preferable to examine the sign of Noether charges. We now have
\begin{align}
&\bar{Q}_a>0:~~q^{(a)}>0,~Q_a>0~~\textrm{or}~~q^{(a)}<0,~Q_a<0,\nonumber \\
&\bar{Q}_a<0:~~q^{(a)}>0,~Q_a<0~~\textrm{or}~~q^{(a)}<0,~Q_a>0,~~~a=1,2.
\end{align}

Without loss of generality, we can select $q^{(a)}\gtrless 0$ for $\bar{Q}_a\gtrless 0$. 
We choose parameters $\omega_a$ such that they satisfy the following conditions 
\be
\begin{aligned}
&\omega_a>q^{(a)}A_t(r)~\textrm{for}~\bar{Q}_a>0\\
&\omega_a<q^{(a)}A_t(r)~\textrm{for}~\bar{Q}_a<0.
\end{aligned} 
\ee

\subsection{Symmetry of equations}
Before we provide the solutions, we comment on an important symmetry of equations. This symmetry combines a gauge symmetry that preserves the electric field $\vec E=-\nabla A_t$ with a transformation that changes the values of the parameters $\omega_a$. The electrostatic field is unaffected by any global change in electrostatic potential $A_t$, i.e. $A_t\rightarrow A_t+C_0$, where $C_0$ is a constant value. This transformation bocames a symmetry of the Poisson equation if the electric charge density remains constant. In general, this is not 
possible until other parameters are replaced. Looking at the form of the function $\rho(r)$ and the formulas \eqref{em:Sigma-f} and \eqref{em:Sigma-g}, we can see that $A_t$ enters these formulas via the functions $b_a(r)=\omega_a-eq^{(a)}A_t(r)$. Thus, if the transformation does not modify the functions $b_a(r),$ the field equations \eqref{em:eq1f}-\eqref{em:eq1a} remain intact. This transformation is of the form
\be
A_t\rightarrow A_t+C_0,\qquad \omega_a\rightarrow\omega_a+e q^{(a)}C_0,\qquad a=1,2.\label{symmetry2}
\ee

Consider replacing $A_t\rightarrow A_t+C_0$ while retaining the values of the parameters $\omega_a$. The functions $b_a$ take the form 
\be
b_a\rightarrow \widetilde b_a=\widetilde \omega_a-eq^{(a)}A_t,\qquad \widetilde\omega_a\equiv\omega_a-eq^{(a)}C_0,\label{gaugetransf}
\ee
which is the same as changing the values of the parameters $\omega_a$. As a result, the size of the compactons changes. They grow when $|\widetilde \omega_a|<|\omega_a|$ and shrink when $|\widetilde \omega_a|>|\omega_a|$. The adjustment of radial profile functions results in a change in charge density, which has an effect on the value of electric charges.

\section{Solutions}
\label{sec:sec3}

In this section, we provide our numerical results for ball-shell and shell-shell harbor-type solutions. Because of the compact character of the scalar field configuration associated with each $CP^N$ component, we can achieve a solution in the multicomponent model in the form of 
nonoverlapping energy density peaks. The attractive or repulsive electrostatic force  between electrically charged compactons changes their radial dimensions. The section is devoted to an examination of this problem.

\subsection{$Q$-ball -- $Q$-shell solution}
We begin by considering the problem of a $Q$-ball surrounded by a $Q$-shell. The size of the compactons is dependent on the number of scalar fields, $u_m$, that constitute the $CP^N$ field. In configurations where the $Q$-shell and $Q$-ball supports overlap (as shown for certain close values of $N_1$ and $N_2$ in \cite{Klimas:2021eue}), the inclusion of the electromagnetic field introduces additional complications. To simplify our analysis, we select values for the number of scalar fields that guarantee their supports do not overlap, specifically choosing $N_1=3$ for the $Q$-shell and $N_2=23$ for the $Q$-ball. While other nonoverlapping values could be chosen, a comprehensive analysis of all such cases, which do not substantially differ from one another, is beyond the scope of this paper. Crucially, since we are analyzing nonoverlapping compactons (whose only interaction is mediated by the electromagnetic field), the specific numerical values of the parameters $\alpha$, $\beta$, and $\lambda$ appearing in 
Eqs. $\eqref{intlagr}$ and $\eqref{Wint}$ do not need to be fixed for the subsequent analysis.

To obtain a solution, we first solve the field $CP^3$ equation together with the Poisson equation using  the shooting method. The shooting parameters are obtained from a series expansion near $r=0$, namely,  $f(r)=a_1 r+a_2r^2+\ldots$ and $A_t(r)=\alpha_0+\alpha_1 r +\alpha_2 r^2+\ldots$. Then, using the field equations we find 
\begin{align}
f(r)&=a_1 r+\frac{\mu_1^2}{32}r^2+\frac{a_1}{10}\left(2a_1^2-\Big(\omega_1-eq^{(1)}\alpha_0\Big)^2\right) r^3\ldots\label{expansionf}\\
A_t(r)&=\alpha_0-\frac{2}{5}M_1^2eq^{(1)}a_1^2\Big(\omega_1-eq^{(1)}\alpha_0\Big)r^4+\ldots\nonumber
\end{align}
where $a_1$ and $\alpha_0$ are two free parameters.
To achieve the $Q$-ball solution, the values of $\alpha_0$ and $a_1$ are fine-tuned. We first fix $\alpha_0$ and then find a value of $a_1$ such that for a given $r=R^{\rm (out)}_1$, the profile function $f(r)$ and its radial derivative $f'(r)$ vanish. The parameter $\alpha_0$ affects the effective $Q$-ball parameters and, as a result, profile functions, charge density associated with them, and, finally, the overall charge of a 
$Q$-ball $\bar Q_1$. Once $a_0$ is chosen, the shooting method is used to obtain the profile function $f(r)$. Because the $CP^N$ $Q$-balls 
always have a minimum value of $|\omega|$, not all $\alpha_0$ values are acceptable because some $\alpha_0$ values  can effectively lead to small values of $|\widetilde \omega|$. In the first three examples $\alpha_0=0.5$.

\begin{figure}[h!]
\centering
\subfigure[]{\includegraphics[width=0.45\textwidth, height=0.25\textwidth,bb=0 0 550 350]{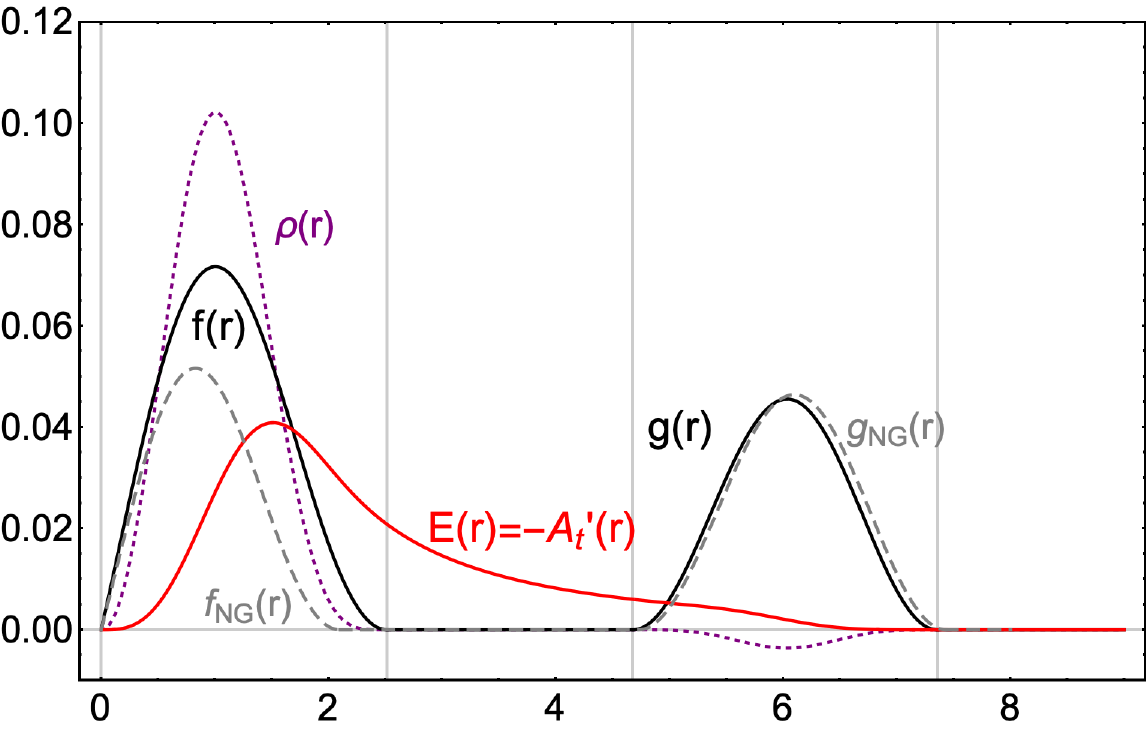}}\hskip0.5cm
\subfigure[]{\includegraphics[width=0.45\textwidth,height=0.25\textwidth,bb=0 0 550 350]{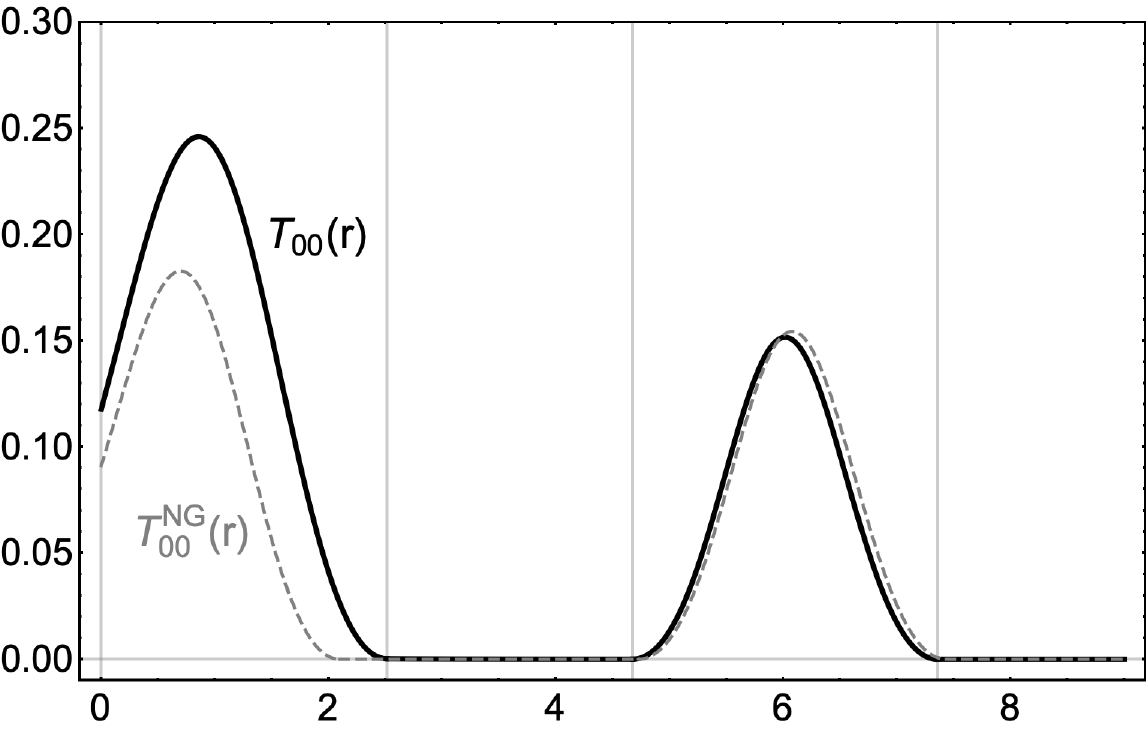}} 
\caption{\label{zeroCP3CP23} The $CP^3$--$CP^{23}$ $Q$-ball -- $Q$-shell configuration with {\it zero} net electric charge $\bar Q_1+\bar Q_2=0$. 
(a) Radial profile functions $f(r)$, $g(r)$ (gauged case -- solid lines; nongauged case -- dashed lines), electric field $E(r)$ 
and electric charge density $\rho (r)$ (dotted line). 
(b) Energy density for gauged (solid line) and nongauged (dashed line) cases.
We employ $(\alpha,\beta,\lambda)=(1.0, 1.0, 1.0)$ for the numerical analysis in this paper, 
but qualitatively, it is not necessary to fix the specific values. }
\end{figure}

The first example, displayed in Fig.\ref{zeroCP3CP23}, depicts the harbor solution, which consists of a positively charged $Q$-ball 
and a negatively charged $Q$-shell.
The $CP^{23}$ field takes its vacuum value $g(r)=0$ in the entire region occupied by the $CP^{3}$ field. The profile function $f(r)$ and the scalar electromagnetic potential $A_t(r)$ are obtained by solving Eqs. \eqref{em:eq1f} and \eqref{em:eq1a}. In this example $\alpha_0=0.5.$  Both scalar field functions have zero values in the region between the outer radius of the $Q$-ball $R_1^{\rm (out)}=2.504$ and the inner radius of the $Q$-shell $R_2^{\rm (in)}=4.666$, and the electrostatic potential defines the Coulomb field of electric charge $\bar Q_1=1.624$. This formula holds true until the internal radius of the $Q$-shell compacton is reached. Then, on the external compacton bulk, the electromagnetic potential is a solution of coupled equations \eqref{em:eq1g} and \eqref{em:eq1a}. Because the $Q$-ball profile function is no longer present, the $CP^{23}$ scalar field is only coupled to the electromagnetic field. The electromagnetic field vanishes beyond the outer $Q$-shell radius $R_2^{\rm (out)}=7.341$ as a consequence of the choice $q^{(2)}=-0.071$, which gives $Q_2=-Q_1$. 

This paper specifically explores scalar field configurations characterized by a zero total net electric charge, where the charges of the constituent components exactly cancel: $Q_1+Q_2=0$.While this is not a generic characteristic of the model, the remarkable property is the very existence of such a possibility. A $Q$-ball-$Q$-shell configuration possessing a total net electric charge exhibits no Coulomb electric field proportional to $r^{-2}$ in the outer region. This situation is analogous to the electric field contained inside a spherical capacitor. However, unlike a simple capacitor, the electric field generated by the inner $Q$-ball influences the charge distribution and, consequently, the size of the external $Q$-shell.

Figure \ref{zeroCP3CP23}(a) depicts the profiles of the radial functions $f(r)$, $g(r)$, and the electric field $E(r)=-A'_t(r)$. The electrostatic repulsion causes the $Q$-ball to inflate, resulting in a larger radius than a neutral $Q$-ball (as seen by the dashed line).
The external $Q$-shell, on the other hand, remains essentially intact because its own electrostatic repulsion is partially compensated by an attraction induced by a $Q$-ball with the opposite charge. At $r=R_2^{\rm (out)}$, the electric field reaches zero value.  The energy density of gauged and nongauged configurations is shown in the final figure, Fig.\ref{zeroCP3CP23}(b).  The contribution of the electric field energy density to the total energy density is quite minimal. The most significant difference is generated by a change in the profiles of the profile functions associated with electrostatic repulsion. When we look at integrals that represent energy, we see that the energy contribution from a scalar field has the value $E_{\rm scalar}=88.147$, but the contribution from an electric field has the value $E_{\rm electric}=0.055$, indicating that the electric field energy is responsible for $0.062\%$ of the total energy.

\begin{figure}[h!]
\centering
\subfigure[]{\includegraphics[width=0.45\textwidth, height=0.25\textwidth,bb=0 0 550 349]{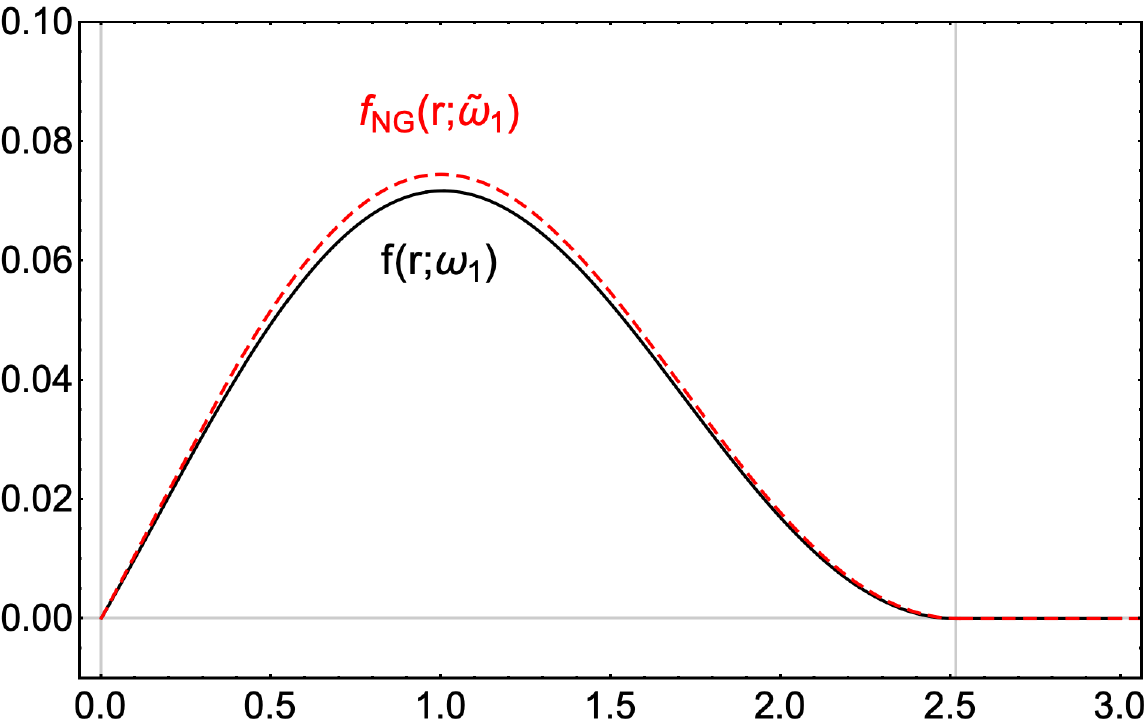}}\hskip0.5cm
\subfigure[]{\includegraphics[width=0.45\textwidth, height=0.25\textwidth,bb=0 0 550 334]{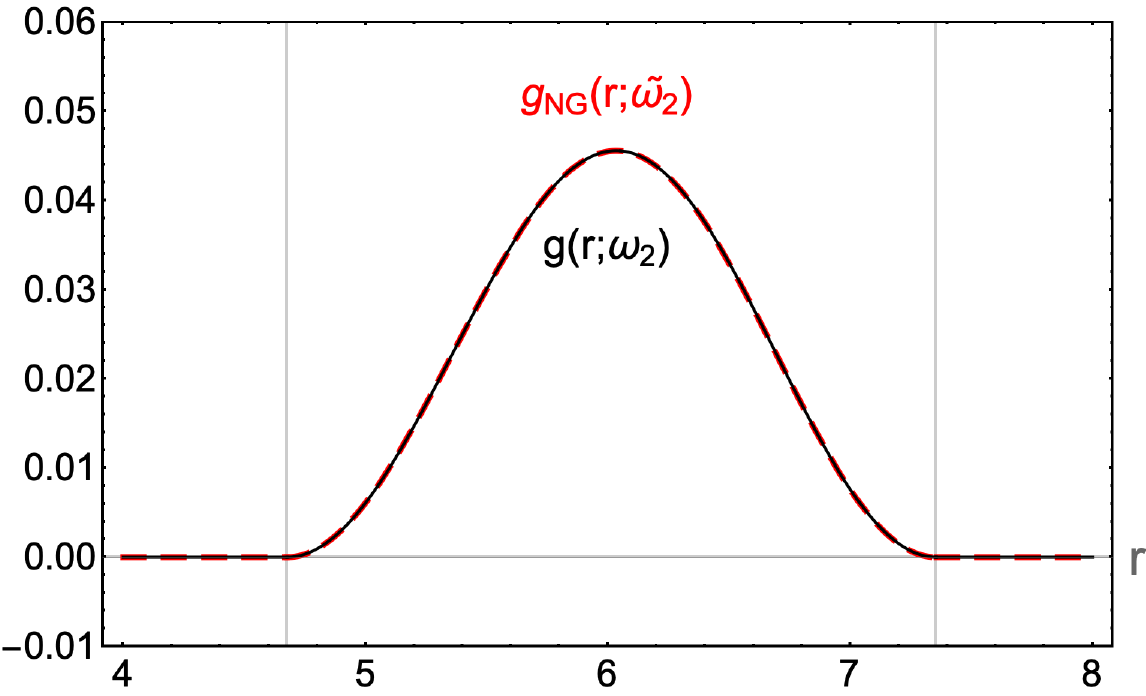}}\hskip0.5cm
\caption{\label{omegarescaled} (a) Profile function $f(r)$ for the gauged model (solid line) parametrized by $\omega_1=3.0$ and $q^{(1)}=1.0$  and the non-gauged profile function $f_{NG}(r)$ (dashed line) for $\widetilde \omega_1=\omega_1-eq^{(1)}\alpha_0=2.5$ where $\alpha_0=0.5$. 
(b) Profile function $g(r)$ for the gauged model (solid line) parametrized by $\omega_1=3.0$ and $q^{(2)}=-0.071$  and the nongauged profile function $g_{NG}(r)$ (dashed line) for $\widetilde \omega_2=\omega_2-eq^{(2)}A_t(R_2^{\rm(out)})\approx 3.029$ where $A_t(R_2^{\rm(out)})=0.418$. }
\end{figure}

The modification in the function form $f(r)$ regarding the electrostatic repulsion can be understood through the changes of the $Q$-ball 
speed $\omega_1\rightarrow \widetilde\omega_1$. In Fig.\ref{omegarescaled}(a), we again draw the profile function $f(r)$ for a gauged $Q$-ball
 and also a profile function $f_{NG}(r)$ for a nongauged $Q$-ball that
marginally includes the effect of $A_t(r)$ by the leading order of the expansion \eqref{expansionf}, 
i.e. $\widetilde\omega_1\equiv\omega_1-eq^{(1)}\alpha_0$ .  
Both profiles are strikingly similar. Therefore, the change is owing to the fact that in the gauged case, 
$A_t$ is not a constant function, and its variation influences the shape of the $Q$-ball function $f(r)$.
We can also understand a very minor change in the $Q$-shell solution by looking at the formula $\widetilde\omega_2=\omega_2-e q^{(2)}A_t(R_2^{\rm(in)})\approx 3.029$. As a result, the profile function $g(r)$ for the gauge $Q$-shell is quite similar to the 
profile function $g_{NG}(r)$ for the nongauged $Q$-shell for $\omega_2=3.0$. The $Q$-shell profiles are shown in Fig.\ref{omegarescaled}(b).

\begin{figure}[h!]
\centering
\subfigure[]{\includegraphics[width=0.45\textwidth, height=0.25\textwidth,bb=0 0 550 334]{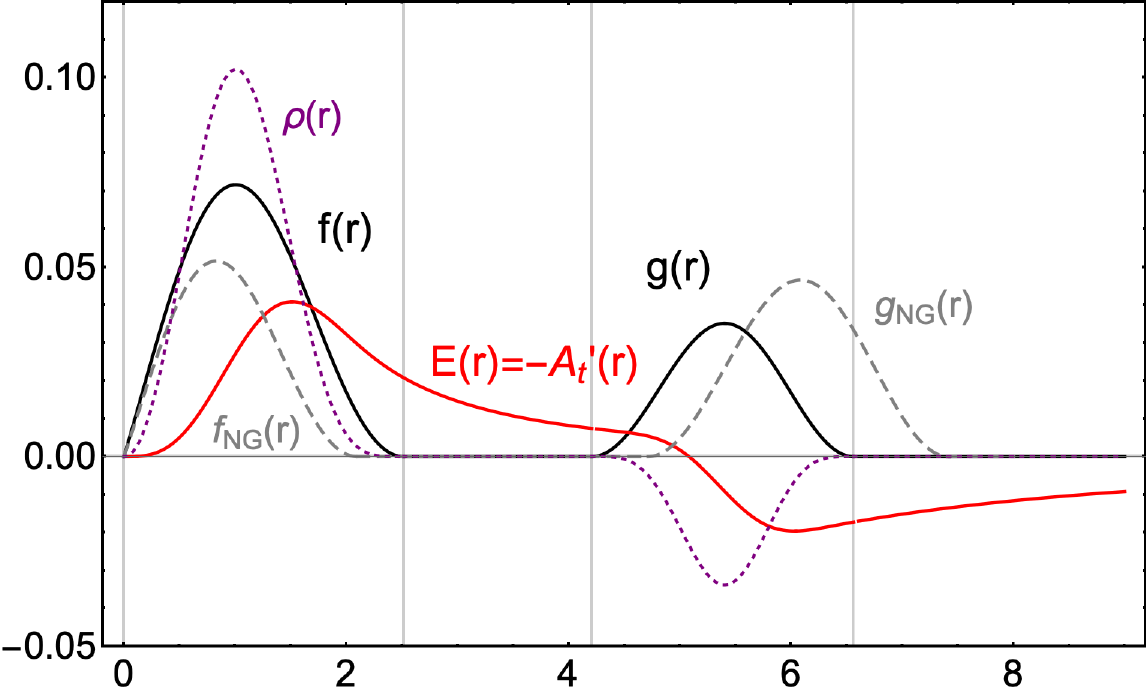}}\hskip0.5cm
\subfigure[]{\includegraphics[width=0.45\textwidth,height=0.25\textwidth,bb=0 0 550 350]{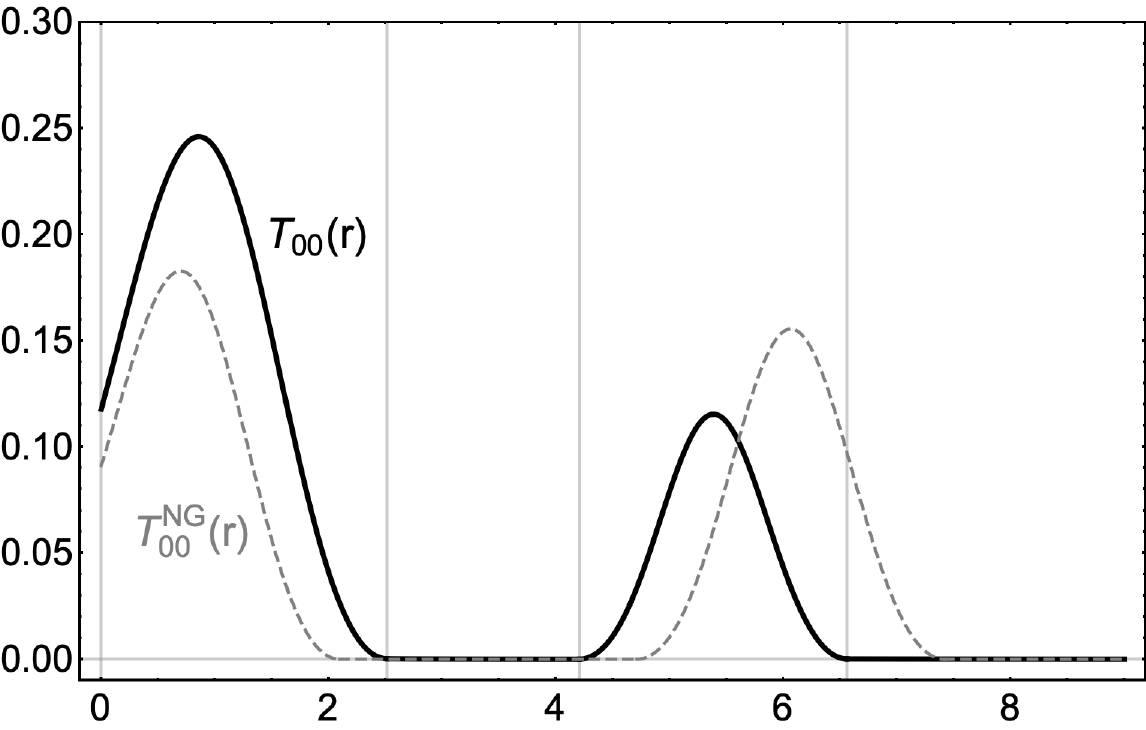}} 
\caption{\label{negativeCP3CP23} The $CP^3$--$CP^{23}$ $Q$-ball -- $Q$-shell configuration with {\it negative} net electric charge $\bar Q_1+\bar Q_2<0$. (a) Radial profile functions $f(r)$, $g(r)$ (gauged case -- solid lines; nongauged case -- dashed lines), the electric field $E(r)$, and the electric charge density (dotted line). (b) Energy density for gauged (solid line) and nongauged (dashed line) cases.}
\end{figure}

It will be interesting to see what happens when the negative charge of the $Q$-shell is increased. We set $q^{(2)}=-1$. The solution in $r\le R_1^{\rm (in)}$ is identical to the previous example, but the value of $R_2^{\rm (in)}=4.200$ is affected by $q^{(2)}$ because it alters the overall value of negative charge. The total net charge reads $\bar Q_1+\bar Q_2=1.624-10.878=-9.253$. As a result, an electric field exists in the region $r>R^{\rm (out)}_2$, where $R^{\rm (out)}_2=6.554$. We present radial profile functions, the electric field and associated scalar potential, the charge density, and the energy density in Fig.\ref{negativeCP3CP23}. The attractive force between $Q$-ball and $Q$-shell increases as the negative charge value increases. As a result, the shell is substantially tighter (lower values of $R_2^{\rm (in)}$ as well as $R_2^{\rm (out)}$ when compared to the nongauged solution). We see that while the gauged $f(r)$ peak is higher than the nongauged one, the $g(r)$ peak is exactly opposite. 

\begin{figure}[h!]
\centering
\subfigure[]{\includegraphics[width=0.45\textwidth, height=0.25\textwidth,bb=0 0 550 350]{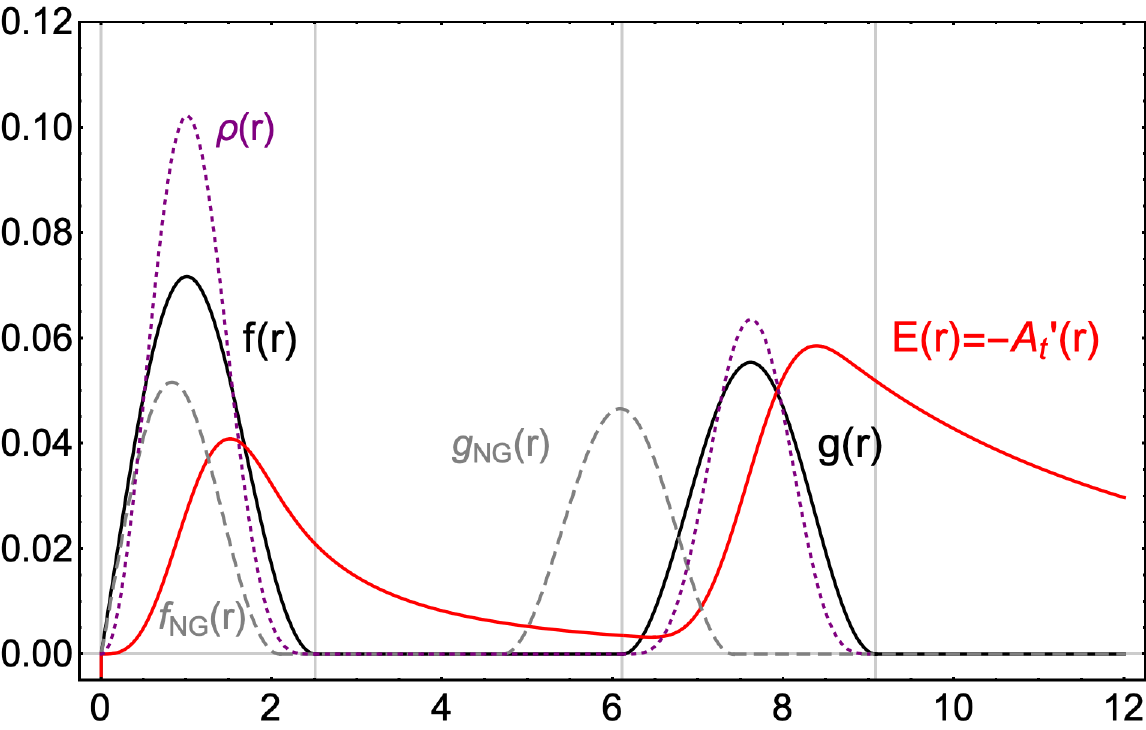}}\hskip0.5cm
\subfigure[]{\includegraphics[width=0.45\textwidth,height=0.25\textwidth,bb=0 0 550 350]{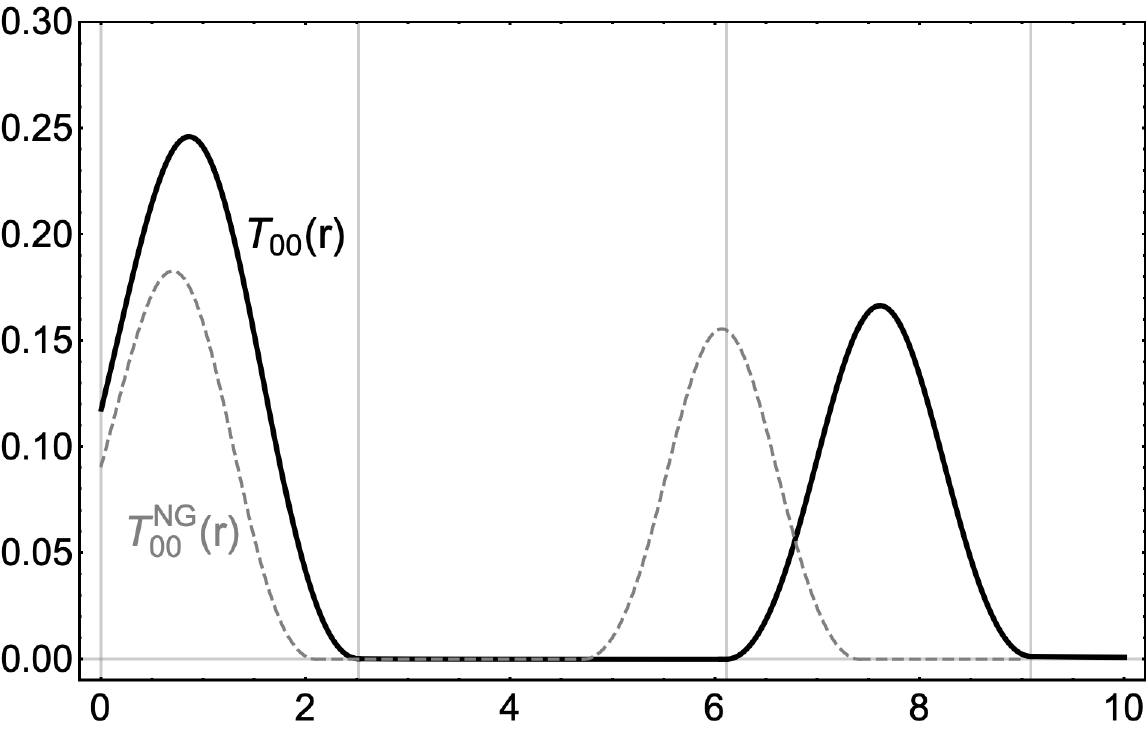}} 
\caption{\label{positiveCP3CP23} The $CP^3$--$CP^{23}$ $Q$-ball -- $Q$-shell configuration with {\it positive} net electric charge $\bar Q_1+\bar Q_2>0$. (a) Radial profile functions $f(r)$, $g(r)$ (gauged case -- solid lines; nongauged case -- dashed lines), the electric field $E(r)$, and the electric charge density (dotted line). (b) Energy density for gauged (solid line) and nongauged (dashed line) cases.}
\end{figure}

Next, we consider a case where $q^{(2)}=+1$, indicating that the $Q$-shell is positively charged. The results are shown in Fig.\ref{positiveCP3CP23}. The repulsive force between charges $\bar Q_1=1.624$ and $\bar Q_2=51.605$ increases radii $R_2^{\rm (in)}$ and $R_2^{\rm (out)}$ which take values $R_2^{\rm (in)}=6.085$ and $R_2^{\rm (out)}=9.056$. Similarly to the previous two examples, $Q$-shell has no effect on the $Q$-ball solution. The electric field of a $Q$-shell in its spherical cavity is zero, as shown by electric Gauss law.

\begin{figure}[h!]
\centering
\subfigure[]{\includegraphics[width=0.45\textwidth, height=0.25\textwidth,bb=0 0 550 341]{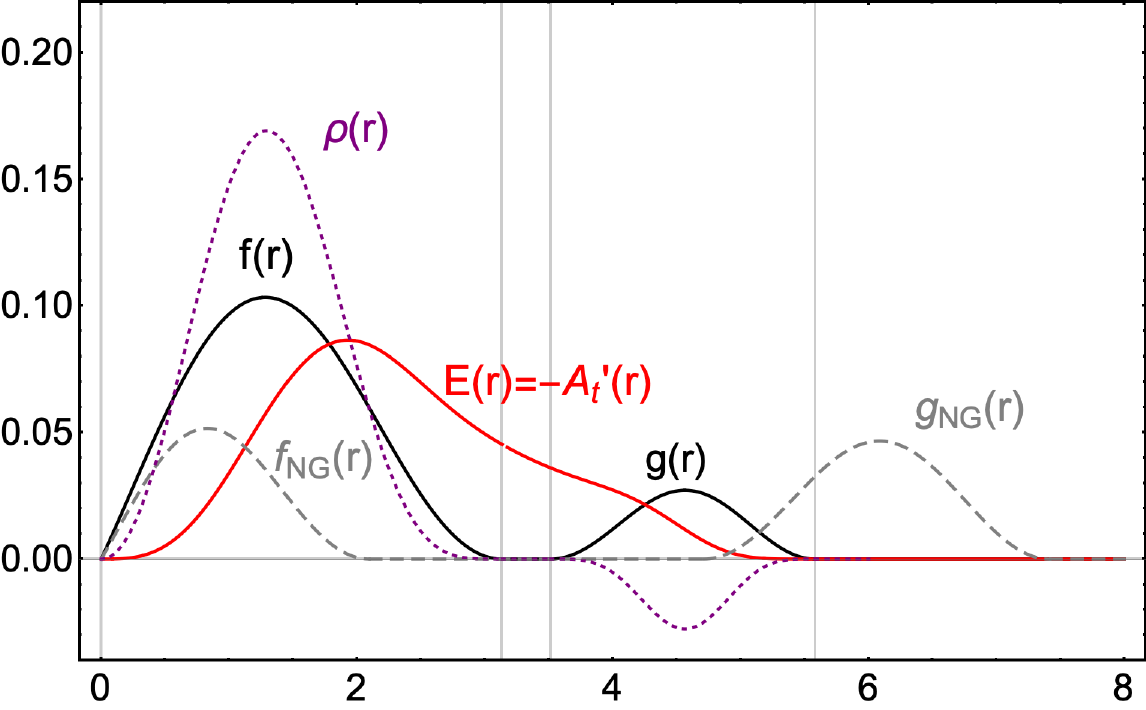}}\hskip0.5cm
\subfigure[]{\includegraphics[width=0.45\textwidth,height=0.25\textwidth,bb=0 0 550 341]{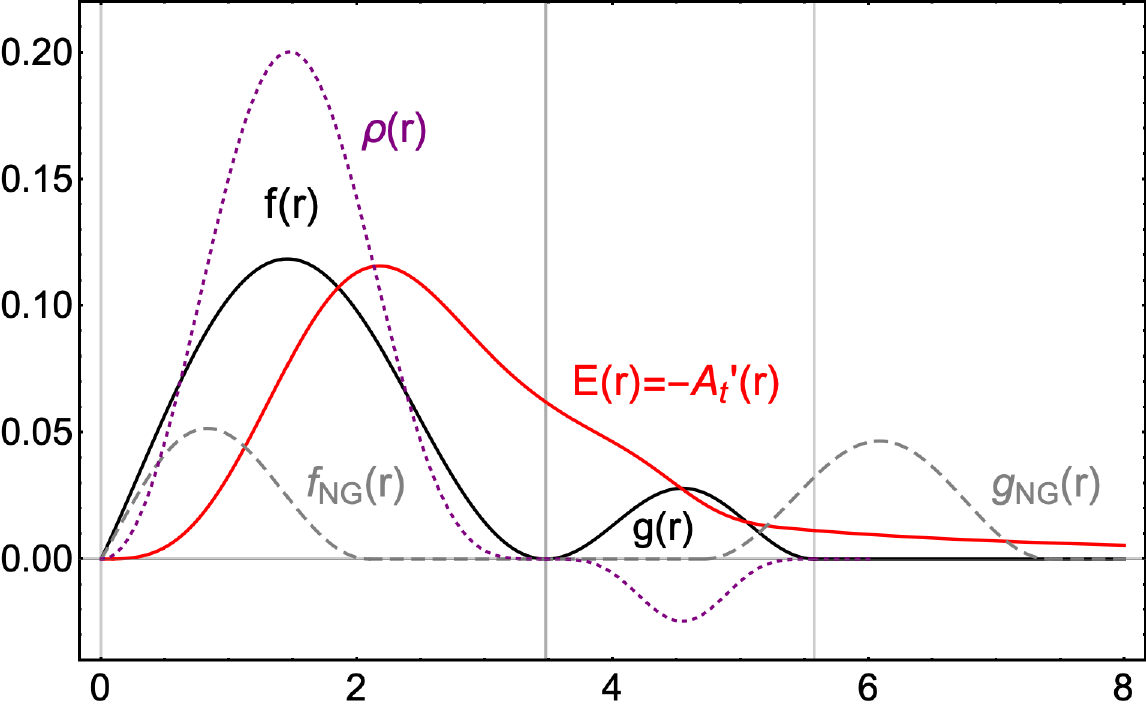}} 
\caption{\label{fig:tight} (a) Zero net charge for $\alpha_0=1.0$ and $q^{(1)}=1.0$. (b) Positive net charge for $\alpha_0=1.2$.}
\end{figure}

Finally, we present two examples of harbor solutions with larger electric charges. In Fig.\ref{fig:tight}(a) 
we plot profile functions for the electric field and electric charge density for solutions with $(\omega_1,\omega_2)=(3.0,3.0)$, 
$\alpha_0=1.0$ and $(q^{(1)},q^{(2)})=(1.0,-1.245)$. The $Q$-ball peak becomes larger than that for nongauged case with $\omega_1=3.0$. 
The gauged $Q$-ball can be approximated with nongauged one having $\widetilde \omega_1=2.0$. Similarly, the gauged $Q$-shell peak approximates the nongauged one, which is parametrized by $\widetilde \omega_2\approx 4.04$ (for $A_t(R_2^{\rm (in)})=0.83$. The $Q$-ball external radius in this example is significantly closer to the $Q$-shell internal radius than in the first example shown in Fig.\ref{zeroCP3CP23}. In this example, the electric charges of $Q$-ball and $Q$-shell are $\bar Q_1=-\bar Q_2=5.39$. If the charges are increased further, the $Q$-ball and $Q$-shell supports will partially overlap. 

In Fig.\ref{fig:tight}(b), we show another instance where both supports are virtually touching $(R_1^{\rm(out)},R_2^{\rm(in)})=(3.43,3.48)$. In this example $(\omega_1,\omega_2)=(3.0,3.0)$, $\alpha_0=1.2$ and $(q^{(1)},q^{(2)})=(1.0,-1.0)$. The electric charges have values $(\bar Q_1,\bar Q_2)=(8.99,-4.94)$.  Gray dashed lines indicate nongauged solutions with $\omega_1$ and $\omega_2$. 
To compare these solutions to nongauged $Q$-ball and $Q$-shell, we use
\begin{align}
&\widetilde \omega_1=\omega_1-e q^{(1)}\alpha_0=1.8,\nonumber\\
&\widetilde \omega_2=\omega_1-e q^{(2)}A_t(R_2^{\rm (out)})\approx 3.97\nonumber
\end{align}
where $R_2^{\rm (out)}=3.48$ corresponds to the gauged case (in a nongauged example, $R_2^{\rm (out)}$ is a shooting parameter with the value $3.55$). 
\begin{figure}[h!]
\centering
\subfigure[]{\includegraphics[width=0.45\textwidth, height=0.25\textwidth,bb=0 0 550 341]{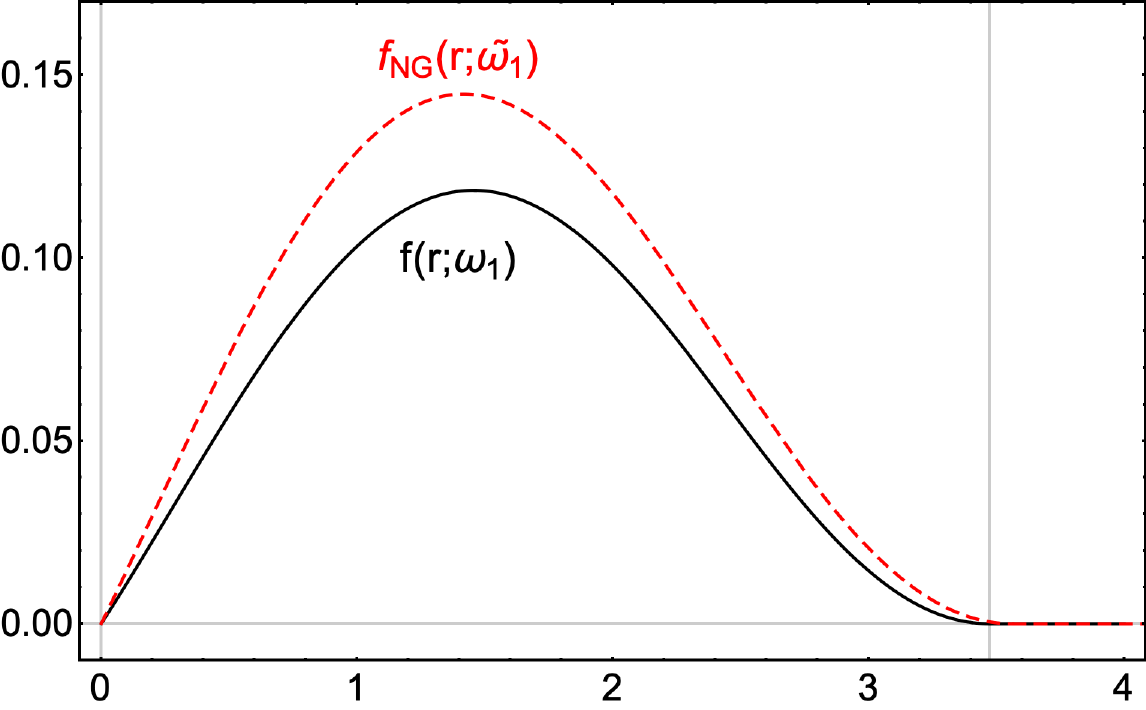}}\hskip0.5cm
\subfigure[]{\includegraphics[width=0.45\textwidth, height=0.25\textwidth,bb=0 0 550 342]{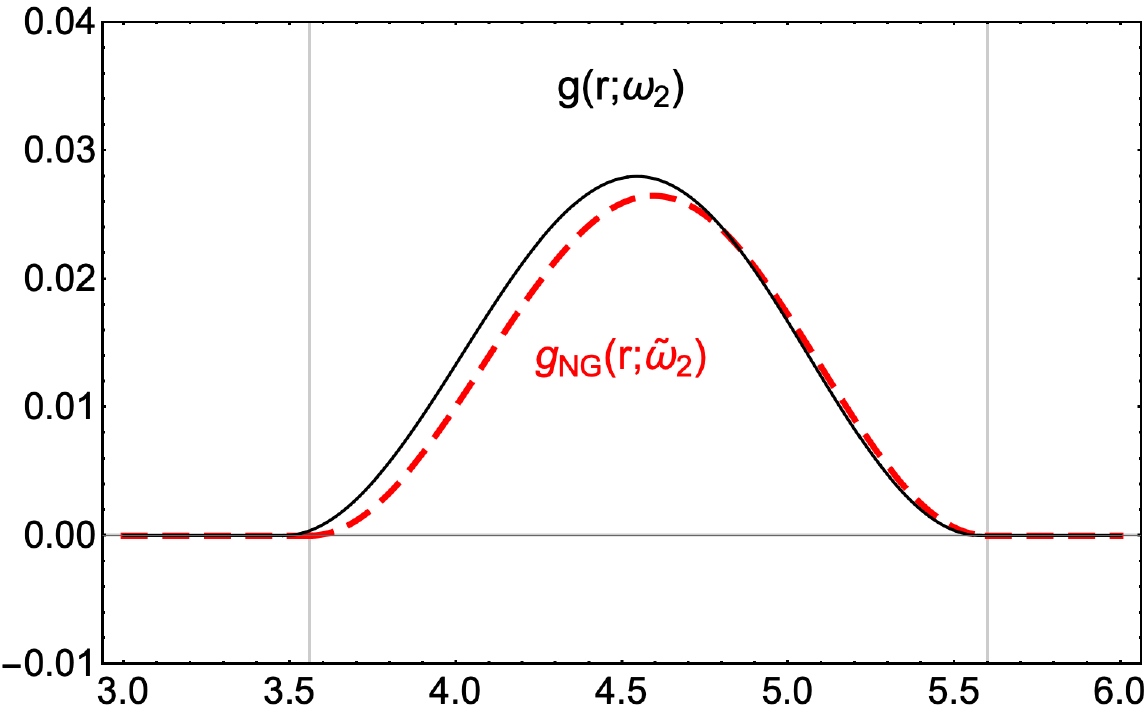}}\hskip0.5cm
\caption{\label{omegarescaled2} (a) Profile function $f(r)$ for gauged model (solid line) parametrized by $\omega_1=3.0$ and $q^{(1)}=1.0$  and the nongauged profile function $f_{NG}(r)$ (dashed line) for $\widetilde \omega_1=\omega_1-eq^{(1)}\alpha_0=1.8$ where $\alpha_0=1.2$. 
(b) Profile function $g(r)$ for gauged model (solid line) parametrized by $\omega_1=3.0$ and $q^{(2)}=-0.071$  and the nongauged profile function $g_{NG}(r)$ (dashed line) for $\widetilde \omega_2=\omega_2-eq^{(2)}A_t(R_2^{\rm(out)})\approx 3.97$ where $A_t(R_2^{\rm(out)})=0.97$. }
\end{figure}
Figure \ref{omegarescaled2} depicts profile functions for gauged and nongauged cases. When compared to Fig.\ref{omegarescaled}, we conclude that, while the shape of the profile functions vary significantly in this situation, the localization of compactons is quite comparable in the gauged and nongauged cases with $\omega_a$ replaced by $\widetilde\omega_a$.

\subsection{$Q$-shell -- $Q$-shell solution}
In this section, we present the harbor solution which consists of two $Q$-shells, one of which is totally within the cavity of the external $Q$-shell.
We explore the behavior of $Q$-shells towards their inner radii  using the shooting method. Because the expansion is quite similar for both $Q$-shells, we display it only once, denoting $(f_1,f_2)=(f,g)$ $R_k\equiv R_k^{\rm(in)}$. Substituting $f_k(r)=\sum_ja_j(r-R_k)^j$ and $A_t(r)=\sum_j\alpha_j(r-R_k)^j$ into field equations yields
\begin{align}
f_k(r)&=\frac{\tilde \mu_k^2}{16}(r-R_k)^2-\frac{\tilde \mu_k^2}{24R}(r-R_k)^3+\nonumber\\
&+\frac{\tilde \mu_k^2}{192R_k^2}\Big[8+l_k(l_k+1)R_k^2(\omega_k-eq^{(k)}\alpha_0)^2\Big](r-R_k)^3+\ldots\label{expshell_f}\\
A_t(r)&=\alpha_0+\alpha_1(r-R_k)-\frac{\alpha_1}{R_k}(r-R_k)^2+\frac{\alpha_1}{R^2_k}(r-R_k)^3-\frac{\alpha_1}{R^3_k}(r-R_k)^4+\ldots\label{expshell_A}
\end{align}
The inner radius of the compact shell $R_k$ plays the role of the shooting parameter. The second derivative $\frac{d^2f_k}{dr^2}|_{r=R_k}$ is fixed by the ratio of $\mu_k^2$ and $M^2_k$. The expansion of the scalar potential, on the other hand, reveals that the value of this potential and its first derivative are free constants. They should be chosen with the current situation in mind. For example, when an external $Q$-shell is exposed to the electric field of the inner $Q$-shell, the central source (charge distribution in the bulk of the central $Q$-shell) determines the electric field and its scalar potential at the inner radius of the external $Q$-shell. The same can be said for the inner $Q$-shell. When the cavity is empty, the scalar potential must be constant inside, implying that $\alpha_1=0$. However, if a pointlike electric charge $\bar Q_{0}$ occurs at $r=0$, it is a source of electric field. As a result, $\alpha_1=-\frac{\bar Q_{0}}{4\pi R_1^2}$. 

\begin{figure}[h!]
\centering
\subfigure[]{\includegraphics[width=0.45\textwidth, height=0.25\textwidth,bb=0 0 550 341]{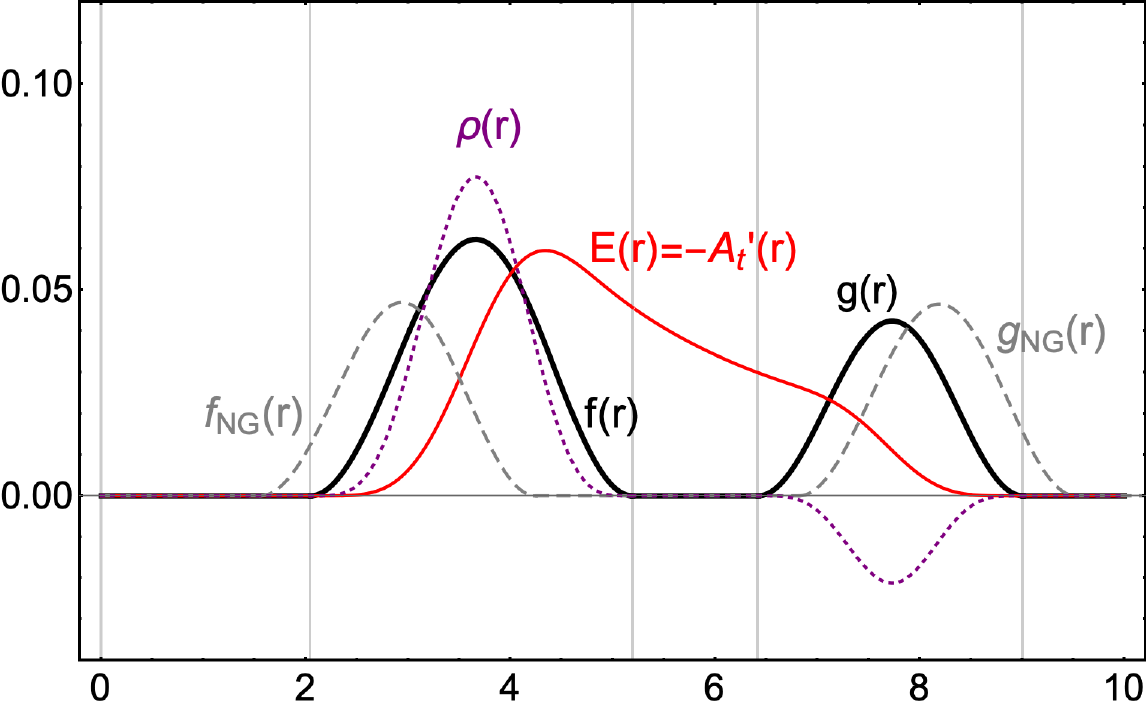}}\hskip0.5cm
\subfigure[]{\includegraphics[width=0.45\textwidth, height=0.25\textwidth,bb=0 0 550 343]{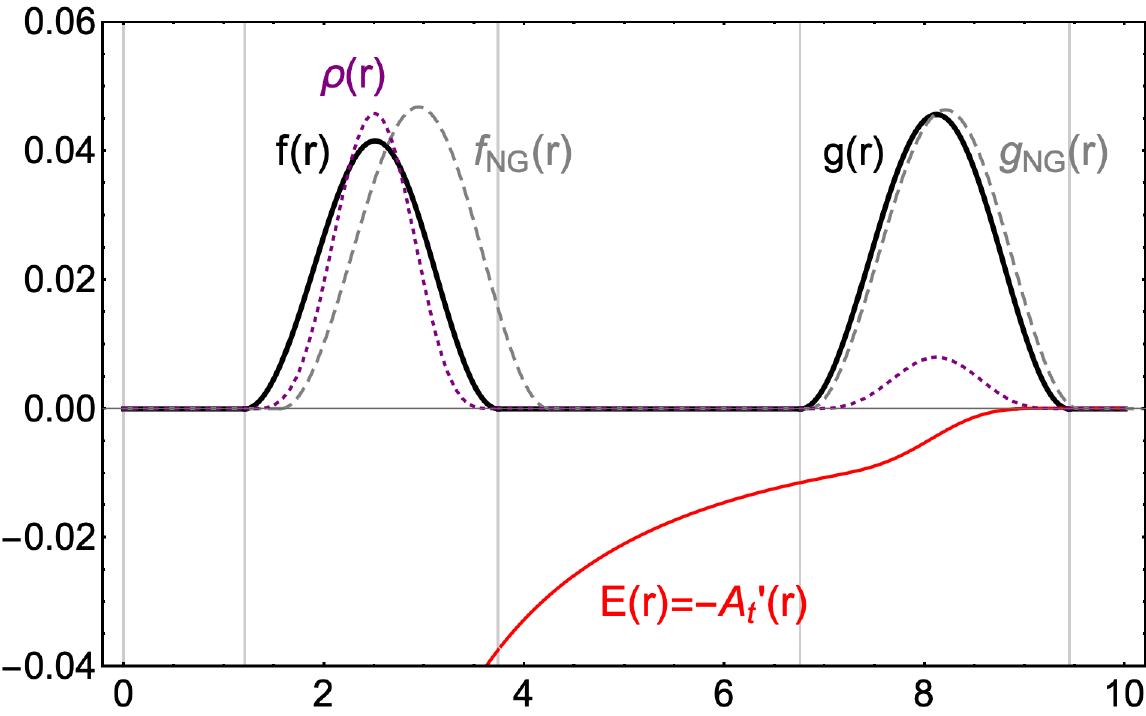}}
\caption{\label{zeroshellshell} The $CP^{11}-CP^{31}$ $Q$-shell -- $Q$-shell configuration with {\it zero} net electric charge $\bar Q_1+\bar Q_2=0$. (a) Radial profile functions $f(r)$, $g(r)$ (gauged case -- solid lines; nongauged case -- dashed lines), the electric field $E(r)$ and electric charge density (dotted line). (b) Another zero net charge configuration $\bar Q_0+\bar Q_1+\bar Q_2=0$ with an extra pointlike charge $\bar Q_{0}=-10.0$ at the center $r=0$.}
\end{figure}
In this example, we present two $Q$-shells in the multicomponent model $CP^{11}-CP^{31}$, i.e., $(l_1,l_2)=(5,15)$. We put $(\omega_1,\omega_2)=(3.0,3.0)$, as in the examples described in the preceding section. We additionally put $\alpha_0=0.5$ and $\alpha_1=0$ at the inner radius of the smallest $Q$-shell because the electric field within the cavity $r<R_1^{\rm(in)}$ is not expected. We solve the field equations for the choice $q^{(1)}=1.0$ and obtain the profile function $f(r)$, the electrostatic potential $A_t(r)$, and the exterior radius of the inner shell. The construction in the region $r>R_1^{\rm(in)}$ is analogous to that of the $Q$-ball--$Q$-shell solution. In this example, we also select $q^{(2)}$ so that the entire electric charge of the field configuration vanishes. The obtained solution is characterized by inner and outer radii of shells $(R_1^{\rm(in)},R_1^{\rm (out)})=(2.032,5.171)$, $(R_2^{\rm(in)},R_2^{\rm (out)})=(6.410,8.996)$, electric charges $(\bar Q_1,\bar Q_2)=(15.26,-15.26)$ and the parameters $(q^{(1)},q^{(2)})=(1.0, -0.464)$. Figure \ref{zeroshellshell}(a) depicts the profile functions, electric field, and charge density. In addition, for nongauged solutions, we draw profile functions with $(\omega_1,\omega_2)=(3.0,3.0)$ (grey dashed line). It is clear that the external $Q$-shell is tighter than in the nongauged case. The attraction to the opposite charge associated with the internal $Q$-shell is stronger than the repulsion induced by the correct external shell's negative charge.

Figure \ref{zeroshellshell}(b) displays the profile functions, electric field, and charge density for another zero net charge configuration in the model $CP^{11}-CP^{31}$ with $(\omega_1,\omega_2)=(3.0,3.0)$. In this case, an extra pointlike electric charge $\bar Q_0=-10.0$ is placed at $r=0$. The electrostatic potential of the pointlike charge at $R_1^{\rm(in)}$ determines parameters of expansion at the inner $Q$-shell radius
\[
\alpha_0=\frac{\bar Q_0}{4\pi R_1^{\rm(in)}},\qquad  \alpha_1=-\frac{\bar Q_0}{4\pi (R_1^{\rm(in)})^2}.
\]
 The inner and outer radii of the $Q$-shell take values $(R_1^{\rm(in)},R_1^{\rm (out)})=(1.207,3.733)$, $(R_2^{\rm(in)},R_2^{\rm (out)})=(6.745,426)$. 
The electric charges of $Q$-shells have values $(\bar Q_1,\bar Q_2)=(3.39(6),6.59(9))\approx(3.40,6.60)$ and 
they correspond to $(q^{(1)},q^{(2)})=(1.0, 0.158)$. 
When the model is coupled to gravity, the harbor solution with pointlike charge can be replaced by a Reissner-Nordstr\"om black hole.

\begin{figure}[h!]
\centering
\includegraphics[width=0.9\linewidth]{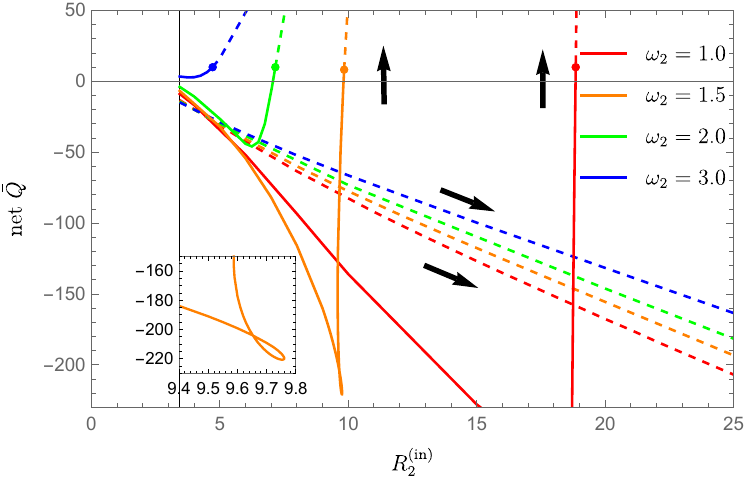}
\caption{\label{phasediagram} 
Phase diagram of the $CP^{3}-CP^{23}$ $Q$-ball -- $Q$-shell. 
The net-zero charge $\bar{Q}$ versus the inner shell $R^{\rm (in)}_2$ is plotted. 
The solid lines indicate the negative charge $q^{(2)}<0$, while the dotted lines are 
the positive charge $q^{(2)}>0$. The point where $q^{(2)}=0$ is shown by the bold dots.
The charge of the ball is fixed by $Q_1=+10$.
The arrows point in the direction 
of increasing $q^{(2)}$.}
\end{figure}

\subsection{Existence of the net-zero solutions of the composite $Q$-ball -- $Q$-shell}

In this paper, we have presented several composite solutions of the $Q$-ball -- $Q$-shell 
and also the $Q$-shell -- $Q$-shell. A particularly nice finding is that the model has a 
solution with zero electric charge, which is recognizable as a neutral object from the outside. 
It is well known that the energy and charge have some scaling relations 
in the basic $Q$-ball model. However, it should be completely rewritten of this type 
solution. In addition, the solution does not always exists; in fact, 
there may be a certain window for such a special solution. In this subsection, we 
numerically examine the parameter condition for the existence of the net-zero solution. 

Figure \ref{phasediagram} presents the net charge $\bar{Q}$ versus the inner radius of 
the shell $R^{\rm (in)}_2$ of the $CP^3-CP^{23}$ $Q$-ball -- $Q$-shell solutions where 
the $Q$-ball charge is fixed as $\bar{Q}_1=+10$. 
We acquire the shell solutions with many charges from several $q^{(2)} (\lesssim 0)$ with 
$\omega_2=1.0-3.0$. 
The behavior can be understood an intuitive way in terms of the following discussion. 
The temporal component of the gauge field $A_t(r)$ 
is assumed to be a positive, monotonically decreasing function without sacrificing generality. 
This means that $b_2(r)$ is always positive for $q^{(2)}<0$, 
as can be directly confirmed by \eqref{ba}. 
As a result, the charge $\bar{Q}_2$ is always negative by the definition~\eqref{Charge2}. 
In Fig.\ref{phasediagram}, we plot the solid lines for the $q^{(2)}<0$ solutions, 
which of course are $\bar{Q}<0$ for $\omega_2 \lesssim 2.0$. 
The dots on the line indicate the solutions with $q^{(2)}=0$. 
The solutions are of $q^{(2)}>0$, which is shown by the 
dotted lines,  where $\bar{Q}_2$ becomes positive because $b_2(r)>0$. 
But when $q^{(2)}$ increases, $b(r)<0$ can exist, and the integral in \eqref{Charge2} 
shifts to the negative. At some large $R_2^{\rm (in)}$, $\bar{Q}_2$ switches to negative 
and then returns to the small $R_2^{\rm (in)}$ region as $q^{(2)}$ decreases
(near the straight dotted lines with $\bar{Q}<0$). 
The arrows in Fig.\ref{phasediagram} show the increase in $q^{(2)}$. 
There are net-zero solutions in $\omega_2 \lesssim 2.0$. 
It is natural to speculate that the far-off $R_2^{\rm (in)}$ region should contain the 
solutions with maximal $q^{(2)}$ as well as the other net-zero solutions. 
Finding such solutions, however, will be a quite challenging task, 
with severe numerical difficulties.

\begin{figure}[t]
\centering
\subfigure[]{\includegraphics[width=0.45\textwidth, height=0.25\textwidth,bb=0 0 550 341]{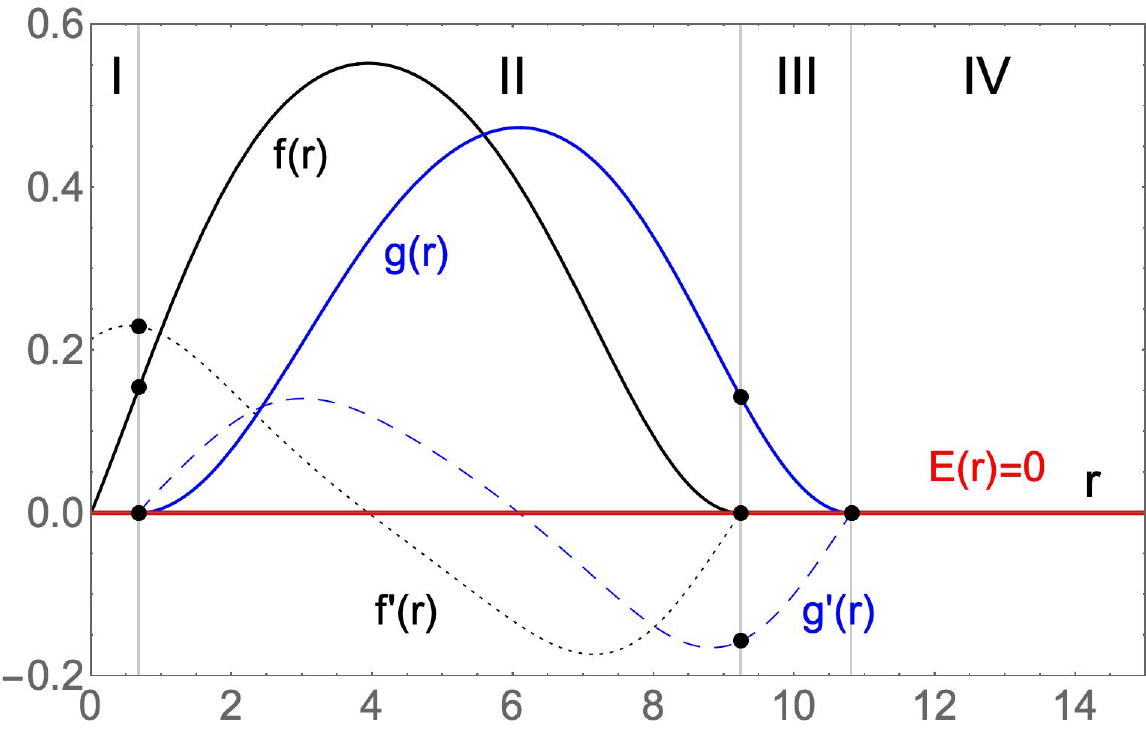}}\hskip0.5cm
\subfigure[]{\includegraphics[width=0.45\textwidth, height=0.25\textwidth,bb=0 0 550 343]{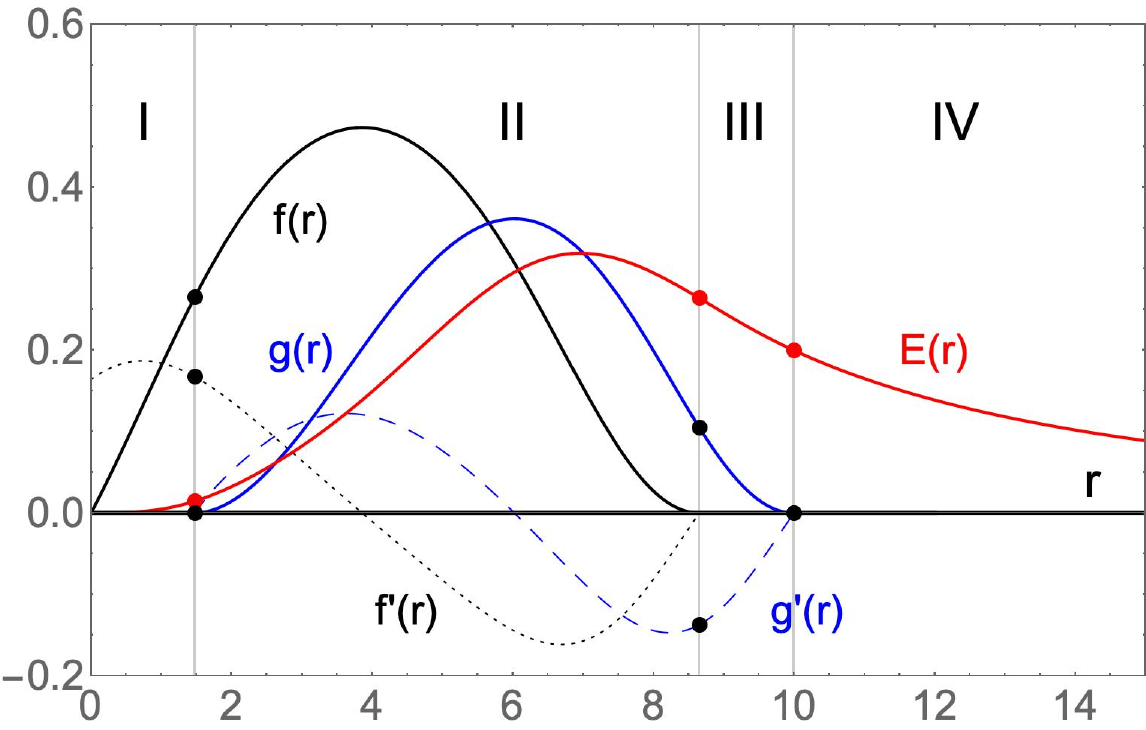}}
\caption{\label{overlap} The $CP^3-CP^5$ overlapping compacton configuration. Panel (a) shows the noncharged compactons, 
where the radii are as follows: $Q$-ball $R^{(\text{out})}=9.240$, and $Q$-shell $R^{(\text{in})}=0.683$, $R^{(\text{out})}=10.817$. Panel (b) shows the electrically charged compactons, where the radii are as follows: $Q$-ball $R^{(\text{out})}=8.649$, and $Q$-shell $R^{(\text{in})}=1.477$, $R^{(\text{out})}=9.997$.}
\end{figure}


\subsection{Overlapping compactons}
Our primary focus in this paper is the study of compact $Q$-balls and $Q$-shells that interact solely through their electric charges. 
However, it is a natural and interesting extension to explore whether gauged compactons can also interact via the potential term. 
In fact, similar to the nongauged case, such overlapping solutions are possible. For future analysis, studying such overlapping 
compactons--for instance, in a $CP^3$--$CP^5$ model--would allow for an interaction that is a combination of electromagnetic effects and terms mediated by the potential. Analyzing these complex scalar field configurations requires solving the system of coupled radial equations in distinct, nonzero field domains. To achieve this, the size of these domains (related to the inner and outer radii of the compactons) must be carefully fine-tuned to ensure the desired quadratic behavior at the compacton boundary. Such a systematic study is essential to thoroughly explore the wide range of possibilities by fixing the form of the interaction potential.

In Fig. \ref{overlap}, we present an example illustrating an overlapping compacton solution featuring two distinct scenarios. Panel (a) shows the interaction between a $CP^3$ $Q$-ball and a $CP^5$ $Q$-shell where both constituents have zero electric charge ($q^{(1)}=q^{(2)}=0$). Panel (b) illustrates the case with the presence of an electric field, obtained by choosing the equal nonzero charges $q^{(1)}=q^{(2)}=0.1$. For both scenarios, the interaction coupling constants in Eqs. \eqref{em:eq1f} and \eqref{em:eq1g} were set as $\lambda_1=\lambda_2=1.0$ and $\alpha=\beta=1.0$. The core system parameters were also fixed: $\omega_1=\omega_2=2.0$ and $\tilde \mu_1=\tilde \mu_2=1.0$.

The presence of electric charges significantly modifies the compacton geometry. It causes the inner $Q$-shell radius to increase, while, on the other hand, the outer $Q$-shell radius shrinks. Furthermore, the peak amplitude of the profile function $g(r)$ is smaller when electric charges are present compared to the noncharged case.The detailed study of the combination of the self-interaction of scalar fields and the interaction 
of the electromagnetic field is a very interesting and complex question that, in our opinion, deserves a thorough analysis. Especially in this context, it would be crucial to obtain the energy-Noether charge relation as this serves as an important indicator of the stability of the $Q$-ball/$Q$-shell field configurations.

\section{Conclusions and Summary}
\label{sec:sec4}

In this paper, we discussed new $Q$-ball configurations in two coupled gauged $CP^N$ models with nonanalytic potentials.
The presence of a vacuum hole inside compact $Q$-shells enables the construction of 
harbor-type solutions, i.e. compactons surrounded by compactons.
The sign of the electric charge is more important in this kind of configuration than in isolated solutions because there is a possibility 
that the overall charge of the ball--shell configuration vanishes even though their respective charges are not zero. 
As a result, the well-known scaling condition for the $Q$-ball $E\sim |Q|^\alpha,\alpha<1$ may require improvement.
In other words, there are many different $Q$-balls (as well as Q-ball and Q-shell configurations) with the same net charge. 
It is one of the novelties of this work.
We obtained $CP^3$--$CP^{23}$ $Q$-ball -- $Q$-shell configurations and also
$CP^{11}$--$CP^{31}$ $Q$-shell -- $Q$-shell configurations with both positive and negative net charge. 
We gave a detailed discussion of the structures of the resulting configurations. 
The electrostatic interaction between oppositely charged compactons in our multicomponent model brought the constituents closer together. 
The size of two zero net-charge multicompactons varies, with one made of an electrically neutral Q-ball and Q-shell 
and the other comprising compactons with opposing charges. 
Neither of these structures has an electric field at spatial infinity from the viewpoint of an external observer. 
This effect may be crucial in the study of boson star sizes.

We also noticed various solutions. One is that when we fix the charge of the inner body, there exist several 
outer shell solutions with the same net total charge. 
Another difficulty is the existence of the composite's excited state. For the net-zero charged solution 
$(Q_1,Q_2)=(+10, -10)$ of the $CP^3$--$CP^{23}$ $Q$-ball -- $Q$-shell configuration, we only found solutions with very large radii. 
Apparently, this case represents the solution's excited state. We briefly examined the fact that the ground state is not present in this scenario. 

There are several potential applications for the harbor solutions. 
When the model is coupled to gravity, the harbor solution with a Reissner-Nordstr\"om black hole becomes 
the harboring black-hole solution~\cite{Sawado:2020ncc}. In this paper, we concentrated on 
the two-components model and the solution forming the two shells. 
The multishell solutions with a combination of many charges $Q_1,Q_2,\cdots,Q_n$, 
which are all different solutions, can be obtained if we take into account a larger number of components. 
The solutions possessing similar structures have already been discussed
~\cite{Loginov:2020xoj,Klimas:2022ghu} in somewhat different ways. 
Such Q-combs provide a totally new perspective for the physics of the $Q$-balls. 

Another interesting direction for future study is the analysis of overlapping compacton configurations, 
which would allow for a thorough exploration of the combined self-interaction and electromagnetic 
interaction effects between $Q$-balls and $Q$-shells.

We will address these issues in future studies.

\begin{center}
{\bf Acknowledgment}
\end{center}
The authors would like to thank Filip Blaschke, Sven Bjarke Gudnason, 
Luiz Agostinho Ferreira and Wojtek Zakrzewski for useful advice and comments. 
N.S. and S.Y. would like to thank all the conference organizers of QTS12 and 
Professor \v{C}estmir Burd\'{i}k for the hospitality.  
N.S. was supported in part by JSPS KAKENHI Grants No. JP20K03278, and No. JP23K02794.
L.C.K. was supported by a FAPESC scholarship.

\bibliography{gaugedQballs}

\begin{thebibliography}{38}%
\makeatletter
\providecommand \@ifxundefined [1]{%
 \@ifx{#1\undefined}
}%
\providecommand \@ifnum [1]{%
 \ifnum #1\expandafter \@firstoftwo
 \else \expandafter \@secondoftwo
 \fi
}%
\providecommand \@ifx [1]{%
 \ifx #1\expandafter \@firstoftwo
 \else \expandafter \@secondoftwo
 \fi
}%
\providecommand \natexlab [1]{#1}%
\providecommand \enquote  [1]{``#1''}%
\providecommand \bibnamefont  [1]{#1}%
\providecommand \bibfnamefont [1]{#1}%
\providecommand \citenamefont [1]{#1}%
\providecommand \href@noop [0]{\@secondoftwo}%
\providecommand \href [0]{\begingroup \@sanitize@url \@href}%
\providecommand \@href[1]{\@@startlink{#1}\@@href}%
\providecommand \@@href[1]{\endgroup#1\@@endlink}%
\providecommand \@sanitize@url [0]{\catcode `\\12\catcode `\$12\catcode
  `\&12\catcode `\#12\catcode `\^12\catcode `\_12\catcode `\%12\relax}%
\providecommand \@@startlink[1]{}%
\providecommand \@@endlink[0]{}%
\providecommand \url  [0]{\begingroup\@sanitize@url \@url }%
\providecommand \@url [1]{\endgroup\@href {#1}{\urlprefix }}%
\providecommand \urlprefix  [0]{URL }%
\providecommand \Eprint [0]{\href }%
\providecommand \doibase [0]{http://dx.doi.org/}%
\providecommand \selectlanguage [0]{\@gobble}%
\providecommand \bibinfo  [0]{\@secondoftwo}%
\providecommand \bibfield  [0]{\@secondoftwo}%
\providecommand \translation [1]{[#1]}%
\providecommand \BibitemOpen [0]{}%
\providecommand \bibitemStop [0]{}%
\providecommand \bibitemNoStop [0]{.\EOS\space}%
\providecommand \EOS [0]{\spacefactor3000\relax}%
\providecommand \BibitemShut  [1]{\csname bibitem#1\endcsname}%
\let\auto@bib@innerbib\@empty
\bibitem [{\citenamefont {Lee}\ and\ \citenamefont {Pang}(1992)}]{Lee:1991ax}%
  \BibitemOpen
  \bibfield  {author} {\bibinfo {author} {\bibfnamefont {T.~D.}\ \bibnamefont
  {Lee}}\ and\ \bibinfo {author} {\bibfnamefont {Y.}~\bibnamefont {Pang}},\
  }\bibfield  {title} {\enquote {\bibinfo {title} {{Nontopological
  solitons}},}\ }\href {\doibase 10.1016/0370-1573(92)90064-7} {\bibfield
  {journal} {\bibinfo  {journal} {Phys. Rept.}\ }\textbf {\bibinfo {volume}
  {221}},\ \bibinfo {pages} {251--350} (\bibinfo {year} {1992})}\BibitemShut
  {NoStop}%
\bibitem [{\citenamefont {Coleman}(1985)}]{Coleman:1985ki}%
  \BibitemOpen
  \bibfield  {author} {\bibinfo {author} {\bibfnamefont {Sidney~R.}\
  \bibnamefont {Coleman}},\ }\bibfield  {title} {\enquote {\bibinfo {title}
  {{Q-balls}},}\ }\href {\doibase 10.1016/0550-3213(86)90520-1} {\bibfield
  {journal} {\bibinfo  {journal} {Nucl. Phys. B}\ }\textbf {\bibinfo {volume}
  {262}},\ \bibinfo {pages} {263} (\bibinfo {year} {1985})},\ \bibinfo {note}
  {[Addendum: Nucl.Phys.B 269, 744 (1986)]}\BibitemShut {NoStop}%
\bibitem [{\citenamefont {Friedberg}\ \emph {et~al.}(1987)\citenamefont
  {Friedberg}, \citenamefont {Lee},\ and\ \citenamefont
  {Pang}}]{Friedberg:1986tq}%
  \BibitemOpen
  \bibfield  {author} {\bibinfo {author} {\bibfnamefont {R.}~\bibnamefont
  {Friedberg}}, \bibinfo {author} {\bibfnamefont {T.~D.}\ \bibnamefont {Lee}},
  \ and\ \bibinfo {author} {\bibfnamefont {Y.}~\bibnamefont {Pang}},\
  }\bibfield  {title} {\enquote {\bibinfo {title} {{Scalar Soliton Stars and
  Black Holes}},}\ }\href {\doibase 10.1103/PhysRevD.35.3658} {\bibfield
  {journal} {\bibinfo  {journal} {Phys. Rev. D}\ }\textbf {\bibinfo {volume}
  {35}},\ \bibinfo {pages} {3658} (\bibinfo {year} {1987})}\BibitemShut
  {NoStop}%
\bibitem [{\citenamefont {Jetzer}(1992)}]{Jetzer:1991jr}%
  \BibitemOpen
  \bibfield  {author} {\bibinfo {author} {\bibfnamefont {Philippe}\
  \bibnamefont {Jetzer}},\ }\bibfield  {title} {\enquote {\bibinfo {title}
  {{Boson stars}},}\ }\href {\doibase 10.1016/0370-1573(92)90123-H} {\bibfield
  {journal} {\bibinfo  {journal} {Phys. Rept.}\ }\textbf {\bibinfo {volume}
  {220}},\ \bibinfo {pages} {163--227} (\bibinfo {year} {1992})}\BibitemShut
  {NoStop}%
\bibitem [{\citenamefont {Liebling}\ and\ \citenamefont
  {Palenzuela}(2012)}]{Liebling:2012fv}%
  \BibitemOpen
  \bibfield  {author} {\bibinfo {author} {\bibfnamefont {Steven~L.}\
  \bibnamefont {Liebling}}\ and\ \bibinfo {author} {\bibfnamefont {Carlos}\
  \bibnamefont {Palenzuela}},\ }\bibfield  {title} {\enquote {\bibinfo {title}
  {{Dynamical Boson Stars}},}\ }\href {\doibase 10.12942/lrr-2012-6} {\bibfield
   {journal} {\bibinfo  {journal} {Living Rev. Rel.}\ }\textbf {\bibinfo
  {volume} {15}},\ \bibinfo {pages} {6} (\bibinfo {year} {2012})},\ \Eprint
  {http://arxiv.org/abs/1202.5809} {arXiv:1202.5809 [gr-qc]} \BibitemShut
  {NoStop}%
\bibitem [{\citenamefont {Kleihaus}\ \emph {et~al.}(2009)\citenamefont
  {Kleihaus}, \citenamefont {Kunz}, \citenamefont {Lammerzahl},\ and\
  \citenamefont {List}}]{Kleihaus:2009kr}%
  \BibitemOpen
  \bibfield  {author} {\bibinfo {author} {\bibfnamefont {Burkhard}\
  \bibnamefont {Kleihaus}}, \bibinfo {author} {\bibfnamefont {Jutta}\
  \bibnamefont {Kunz}}, \bibinfo {author} {\bibfnamefont {Claus}\ \bibnamefont
  {Lammerzahl}}, \ and\ \bibinfo {author} {\bibfnamefont {Meike}\ \bibnamefont
  {List}},\ }\bibfield  {title} {\enquote {\bibinfo {title} {{Charged Boson
  Stars and Black Holes}},}\ }\href {\doibase 10.1016/j.physletb.2009.03.066}
  {\bibfield  {journal} {\bibinfo  {journal} {Phys. Lett. B}\ }\textbf
  {\bibinfo {volume} {675}},\ \bibinfo {pages} {102--115} (\bibinfo {year}
  {2009})},\ \Eprint {http://arxiv.org/abs/0902.4799} {arXiv:0902.4799 [gr-qc]}
  \BibitemShut {NoStop}%
\bibitem [{\citenamefont {Kleihaus}\ \emph {et~al.}(2010)\citenamefont
  {Kleihaus}, \citenamefont {Kunz}, \citenamefont {Lammerzahl},\ and\
  \citenamefont {List}}]{Kleihaus:2010ep}%
  \BibitemOpen
  \bibfield  {author} {\bibinfo {author} {\bibfnamefont {Burkhard}\
  \bibnamefont {Kleihaus}}, \bibinfo {author} {\bibfnamefont {Jutta}\
  \bibnamefont {Kunz}}, \bibinfo {author} {\bibfnamefont {Claus}\ \bibnamefont
  {Lammerzahl}}, \ and\ \bibinfo {author} {\bibfnamefont {Meike}\ \bibnamefont
  {List}},\ }\bibfield  {title} {\enquote {\bibinfo {title} {{Boson Shells
  Harbouring Charged Black Holes}},}\ }\href {\doibase
  10.1103/PhysRevD.82.104050} {\bibfield  {journal} {\bibinfo  {journal} {Phys.
  Rev. D}\ }\textbf {\bibinfo {volume} {82}},\ \bibinfo {pages} {104050}
  (\bibinfo {year} {2010})},\ \Eprint {http://arxiv.org/abs/1007.1630}
  {arXiv:1007.1630 [gr-qc]} \BibitemShut {NoStop}%
\bibitem [{\citenamefont {Arodz}\ and\ \citenamefont
  {Lis}(2008)}]{Arodz:2008jk}%
  \BibitemOpen
  \bibfield  {author} {\bibinfo {author} {\bibfnamefont {H.}~\bibnamefont
  {Arodz}}\ and\ \bibinfo {author} {\bibfnamefont {J.}~\bibnamefont {Lis}},\
  }\bibfield  {title} {\enquote {\bibinfo {title} {{Compact Q-balls in the
  complex signum-Gordon model}},}\ }\href {\doibase 10.1103/PhysRevD.77.107702}
  {\bibfield  {journal} {\bibinfo  {journal} {Phys. Rev. D}\ }\textbf {\bibinfo
  {volume} {77}},\ \bibinfo {pages} {107702} (\bibinfo {year} {2008})},\
  \Eprint {http://arxiv.org/abs/0803.1566} {arXiv:0803.1566 [hep-th]}
  \BibitemShut {NoStop}%
\bibitem [{\citenamefont {Rosenau}\ and\ \citenamefont
  {Hyman}(1993)}]{Rosenau:1993zz}%
  \BibitemOpen
  \bibfield  {author} {\bibinfo {author} {\bibfnamefont {Philip}\ \bibnamefont
  {Rosenau}}\ and\ \bibinfo {author} {\bibfnamefont {James~M.}\ \bibnamefont
  {Hyman}},\ }\bibfield  {title} {\enquote {\bibinfo {title} {{Compactons:
  Solitons with finite wavelength}},}\ }\href {\doibase
  10.1103/PhysRevLett.70.564} {\bibfield  {journal} {\bibinfo  {journal} {Phys.
  Rev. Lett.}\ }\textbf {\bibinfo {volume} {70}},\ \bibinfo {pages} {564--567}
  (\bibinfo {year} {1993})}\BibitemShut {NoStop}%
\bibitem [{\citenamefont {Rosenau}(1994)}]{PhysRevLett.73.1737}%
  \BibitemOpen
  \bibfield  {author} {\bibinfo {author} {\bibfnamefont {Philip}\ \bibnamefont
  {Rosenau}},\ }\bibfield  {title} {\enquote {\bibinfo {title} {Nonlinear
  dispersion and compact structures},}\ }\href {\doibase
  10.1103/PhysRevLett.73.1737} {\bibfield  {journal} {\bibinfo  {journal}
  {Phys. Rev. Lett.}\ }\textbf {\bibinfo {volume} {73}},\ \bibinfo {pages}
  {1737--1741} (\bibinfo {year} {1994})}\BibitemShut {NoStop}%
\bibitem [{\citenamefont {Adam}\ \emph {et~al.}(2017)\citenamefont {Adam},
  \citenamefont {Sanchez-Guillen},\ and\ \citenamefont
  {Wereszczynski}}]{Adam:2017pdh}%
  \BibitemOpen
  \bibfield  {author} {\bibinfo {author} {\bibfnamefont {C.}~\bibnamefont
  {Adam}}, \bibinfo {author} {\bibfnamefont {J.}~\bibnamefont
  {Sanchez-Guillen}}, \ and\ \bibinfo {author} {\bibfnamefont {A.}~\bibnamefont
  {Wereszczynski}},\ }\bibfield  {title} {\enquote {\bibinfo {title} {{BPS
  submodels of the Skyrme model}},}\ }\href {\doibase
  10.1016/j.physletb.2017.04.003} {\bibfield  {journal} {\bibinfo  {journal}
  {Phys. Lett. B}\ }\textbf {\bibinfo {volume} {769}},\ \bibinfo {pages}
  {362--367} (\bibinfo {year} {2017})},\ \Eprint
  {http://arxiv.org/abs/1703.05818} {arXiv:1703.05818 [hep-th]} \BibitemShut
  {NoStop}%
\bibitem [{\citenamefont {Adam}\ \emph {et~al.}(2018)\citenamefont {Adam},
  \citenamefont {Foster}, \citenamefont {Krusch},\ and\ \citenamefont
  {Wereszczynski}}]{Adam:2017srx}%
  \BibitemOpen
  \bibfield  {author} {\bibinfo {author} {\bibfnamefont {C.}~\bibnamefont
  {Adam}}, \bibinfo {author} {\bibfnamefont {D.}~\bibnamefont {Foster}},
  \bibinfo {author} {\bibfnamefont {S.}~\bibnamefont {Krusch}}, \ and\ \bibinfo
  {author} {\bibfnamefont {A.}~\bibnamefont {Wereszczynski}},\ }\bibfield
  {title} {\enquote {\bibinfo {title} {{BPS sectors of the Skyrme model and
  their non-BPS extensions}},}\ }\href {\doibase 10.1103/PhysRevD.97.036002}
  {\bibfield  {journal} {\bibinfo  {journal} {Phys. Rev. D}\ }\textbf {\bibinfo
  {volume} {97}},\ \bibinfo {pages} {036002} (\bibinfo {year} {2018})},\
  \Eprint {http://arxiv.org/abs/1709.06583} {arXiv:1709.06583 [hep-th]}
  \BibitemShut {NoStop}%
\bibitem [{\citenamefont {Adam}\ \emph
  {et~al.}(2010{\natexlab{a}})\citenamefont {Adam}, \citenamefont
  {Sanchez-Guillen},\ and\ \citenamefont {Wereszczynski}}]{Adam:2010fg}%
  \BibitemOpen
  \bibfield  {author} {\bibinfo {author} {\bibfnamefont {C.}~\bibnamefont
  {Adam}}, \bibinfo {author} {\bibfnamefont {J.}~\bibnamefont
  {Sanchez-Guillen}}, \ and\ \bibinfo {author} {\bibfnamefont {A.}~\bibnamefont
  {Wereszczynski}},\ }\bibfield  {title} {\enquote {\bibinfo {title} {{A
  Skyrme-type proposal for baryonic matter}},}\ }\href {\doibase
  10.1016/j.physletb.2010.06.025} {\bibfield  {journal} {\bibinfo  {journal}
  {Phys. Lett. B}\ }\textbf {\bibinfo {volume} {691}},\ \bibinfo {pages}
  {105--110} (\bibinfo {year} {2010}{\natexlab{a}})},\ \Eprint
  {http://arxiv.org/abs/1001.4544} {arXiv:1001.4544 [hep-th]} \BibitemShut
  {NoStop}%
\bibitem [{\citenamefont {Adam}\ \emph
  {et~al.}(2010{\natexlab{b}})\citenamefont {Adam}, \citenamefont
  {Sanchez-Guillen},\ and\ \citenamefont {Wereszczynski}}]{Adam:2010ds}%
  \BibitemOpen
  \bibfield  {author} {\bibinfo {author} {\bibfnamefont {C.}~\bibnamefont
  {Adam}}, \bibinfo {author} {\bibfnamefont {J.}~\bibnamefont
  {Sanchez-Guillen}}, \ and\ \bibinfo {author} {\bibfnamefont {A.}~\bibnamefont
  {Wereszczynski}},\ }\bibfield  {title} {\enquote {\bibinfo {title} {{A BPS
  Skyrme model and baryons at large $N_c$}},}\ }\href {\doibase
  10.1103/PhysRevD.82.085015} {\bibfield  {journal} {\bibinfo  {journal} {Phys.
  Rev. D}\ }\textbf {\bibinfo {volume} {82}},\ \bibinfo {pages} {085015}
  (\bibinfo {year} {2010}{\natexlab{b}})},\ \Eprint
  {http://arxiv.org/abs/1007.1567} {arXiv:1007.1567 [hep-th]} \BibitemShut
  {NoStop}%
\bibitem [{\citenamefont {Gisiger}\ and\ \citenamefont
  {Paranjape}(1997)}]{Gisiger:1996vb}%
  \BibitemOpen
  \bibfield  {author} {\bibinfo {author} {\bibfnamefont {T.}~\bibnamefont
  {Gisiger}}\ and\ \bibinfo {author} {\bibfnamefont {Manu~B.}\ \bibnamefont
  {Paranjape}},\ }\bibfield  {title} {\enquote {\bibinfo {title} {{Solitons in
  a baby Skyrme model with invariance under volume / area preserving
  diffeomorphisms}},}\ }\href {\doibase 10.1103/PhysRevD.55.7731} {\bibfield
  {journal} {\bibinfo  {journal} {Phys. Rev. D}\ }\textbf {\bibinfo {volume}
  {55}},\ \bibinfo {pages} {7731--7738} (\bibinfo {year} {1997})},\ \Eprint
  {http://arxiv.org/abs/hep-ph/9606328} {arXiv:hep-ph/9606328} \BibitemShut
  {NoStop}%
\bibitem [{\citenamefont {Arodz}\ and\ \citenamefont
  {Lis}(2009)}]{Arodz:2008nm}%
  \BibitemOpen
  \bibfield  {author} {\bibinfo {author} {\bibfnamefont {H.}~\bibnamefont
  {Arodz}}\ and\ \bibinfo {author} {\bibfnamefont {J.}~\bibnamefont {Lis}},\
  }\bibfield  {title} {\enquote {\bibinfo {title} {{Compact Q-balls and
  Q-shells in a scalar electrodynamics}},}\ }\href {\doibase
  10.1103/PhysRevD.79.045002} {\bibfield  {journal} {\bibinfo  {journal} {Phys.
  Rev. D}\ }\textbf {\bibinfo {volume} {79}},\ \bibinfo {pages} {045002}
  (\bibinfo {year} {2009})},\ \Eprint {http://arxiv.org/abs/0812.3284}
  {arXiv:0812.3284 [hep-th]} \BibitemShut {NoStop}%
\bibitem [{\citenamefont {Klimas}\ and\ \citenamefont
  {Livramento}(2017)}]{Klimas:2017eft}%
  \BibitemOpen
  \bibfield  {author} {\bibinfo {author} {\bibfnamefont {P.}~\bibnamefont
  {Klimas}}\ and\ \bibinfo {author} {\bibfnamefont {L.~R.}\ \bibnamefont
  {Livramento}},\ }\bibfield  {title} {\enquote {\bibinfo {title} {{Compact
  Q-balls and Q-shells in CPN type models}},}\ }\href {\doibase
  10.1103/PhysRevD.96.016001} {\bibfield  {journal} {\bibinfo  {journal} {Phys.
  Rev. D}\ }\textbf {\bibinfo {volume} {96}},\ \bibinfo {pages} {016001}
  (\bibinfo {year} {2017})},\ \Eprint {http://arxiv.org/abs/1704.01132}
  {arXiv:1704.01132 [hep-th]} \BibitemShut {NoStop}%
\bibitem [{\citenamefont {Sawado}\ and\ \citenamefont
  {Yanai}(2020)}]{Sawado:2020ncc}%
  \BibitemOpen
  \bibfield  {author} {\bibinfo {author} {\bibfnamefont {Nobuyuki}\
  \bibnamefont {Sawado}}\ and\ \bibinfo {author} {\bibfnamefont {Shota}\
  \bibnamefont {Yanai}},\ }\bibfield  {title} {\enquote {\bibinfo {title}
  {{Compact, charged boson stars and shells in the $\mathbb{C}P^N$ gravitating
  nonlinear sigma model}},}\ }\href {\doibase 10.1103/PhysRevD.102.045007}
  {\bibfield  {journal} {\bibinfo  {journal} {Phys. Rev. D}\ }\textbf {\bibinfo
  {volume} {102}},\ \bibinfo {pages} {045007} (\bibinfo {year} {2020})},\
  \Eprint {http://arxiv.org/abs/2006.03424} {arXiv:2006.03424 [hep-th]}
  \BibitemShut {NoStop}%
\bibitem [{\citenamefont {Klimas}\ \emph {et~al.}(2021)\citenamefont {Klimas},
  \citenamefont {Kubaski}, \citenamefont {Sawado},\ and\ \citenamefont
  {Yanai}}]{Klimas:2021eue}%
  \BibitemOpen
  \bibfield  {author} {\bibinfo {author} {\bibfnamefont {P.}~\bibnamefont
  {Klimas}}, \bibinfo {author} {\bibfnamefont {L.~C.}\ \bibnamefont {Kubaski}},
  \bibinfo {author} {\bibfnamefont {N.}~\bibnamefont {Sawado}}, \ and\ \bibinfo
  {author} {\bibfnamefont {S.}~\bibnamefont {Yanai}},\ }\bibfield  {title}
  {\enquote {\bibinfo {title} {{Compact Q-balls and Q-shells in a
  multi-component \ensuremath{\mathbb{C}}P$^{N}$ model}},}\ }\href {\doibase
  10.1007/JHEP09(2021)084} {\bibfield  {journal} {\bibinfo  {journal} {JHEP}\
  }\textbf {\bibinfo {volume} {09}},\ \bibinfo {pages} {084} (\bibinfo {year}
  {2021})},\ \Eprint {http://arxiv.org/abs/2107.09831} {arXiv:2107.09831
  [hep-th]} \BibitemShut {NoStop}%
\bibitem [{\citenamefont {Zakrzewski}(1989)}]{Zakrzewski1989LowdimensionalSM}%
  \BibitemOpen
  \bibfield  {author} {\bibinfo {author} {\bibfnamefont {W.~J.}\ \bibnamefont
  {Zakrzewski}},\ }\href@noop {} {\emph {\bibinfo {title} {Low-dimensional
  Sigma Models}}}\ (\bibinfo  {publisher} {Adam Hilger},\ \bibinfo {address}
  {Bristol and Philadelphia},\ \bibinfo {year} {1989})\BibitemShut {NoStop}%
\bibitem [{\citenamefont {Ferreira}\ \emph
  {et~al.}(2011{\natexlab{a}})\citenamefont {Ferreira}, \citenamefont
  {Klimas},\ and\ \citenamefont {Zakrzewski}}]{Ferreira:2011nd}%
  \BibitemOpen
  \bibfield  {author} {\bibinfo {author} {\bibfnamefont {L.~A.}\ \bibnamefont
  {Ferreira}}, \bibinfo {author} {\bibfnamefont {P.}~\bibnamefont {Klimas}}, \
  and\ \bibinfo {author} {\bibfnamefont {W.~J.}\ \bibnamefont {Zakrzewski}},\
  }\bibfield  {title} {\enquote {\bibinfo {title} {{Some (3+1) dimensional
  vortex solutions of the CPN model}},}\ }\href {\doibase
  10.1103/PhysRevD.83.105018} {\bibfield  {journal} {\bibinfo  {journal} {Phys.
  Rev. D}\ }\textbf {\bibinfo {volume} {83}},\ \bibinfo {pages} {105018}
  (\bibinfo {year} {2011}{\natexlab{a}})},\ \Eprint
  {http://arxiv.org/abs/1103.0559} {arXiv:1103.0559 [hep-th]} \BibitemShut
  {NoStop}%
\bibitem [{\citenamefont {Ferreira}\ \emph
  {et~al.}(2011{\natexlab{b}})\citenamefont {Ferreira}, \citenamefont
  {Klimas},\ and\ \citenamefont {Zakrzewski}}]{Ferreira:2011jy}%
  \BibitemOpen
  \bibfield  {author} {\bibinfo {author} {\bibfnamefont {L.~A.}\ \bibnamefont
  {Ferreira}}, \bibinfo {author} {\bibfnamefont {P.}~\bibnamefont {Klimas}}, \
  and\ \bibinfo {author} {\bibfnamefont {W.~J.}\ \bibnamefont {Zakrzewski}},\
  }\bibfield  {title} {\enquote {\bibinfo {title} {{Properties of some (3+1)
  dimensional vortex solutions of the $CP^N$ model}},}\ }\href {\doibase
  10.1103/PhysRevD.84.085022} {\bibfield  {journal} {\bibinfo  {journal} {Phys.
  Rev. D}\ }\textbf {\bibinfo {volume} {84}},\ \bibinfo {pages} {085022}
  (\bibinfo {year} {2011}{\natexlab{b}})},\ \Eprint
  {http://arxiv.org/abs/1108.4401} {arXiv:1108.4401 [hep-th]} \BibitemShut
  {NoStop}%
\bibitem [{\citenamefont {Ferreira}(2009)}]{Ferreira:2008nn}%
  \BibitemOpen
  \bibfield  {author} {\bibinfo {author} {\bibfnamefont {L.~A.}\ \bibnamefont
  {Ferreira}},\ }\bibfield  {title} {\enquote {\bibinfo {title} {{Exact vortex
  solutions in an extended Skyrme-Faddeev model}},}\ }\href {\doibase
  10.1088/1126-6708/2009/05/001} {\bibfield  {journal} {\bibinfo  {journal}
  {JHEP}\ }\textbf {\bibinfo {volume} {05}},\ \bibinfo {pages} {001} (\bibinfo
  {year} {2009})},\ \Eprint {http://arxiv.org/abs/0809.4303} {arXiv:0809.4303
  [hep-th]} \BibitemShut {NoStop}%
\bibitem [{\citenamefont {Ferreira}\ and\ \citenamefont
  {Klimas}(2010)}]{Ferreira:2010jb}%
  \BibitemOpen
  \bibfield  {author} {\bibinfo {author} {\bibfnamefont {L.~A.}\ \bibnamefont
  {Ferreira}}\ and\ \bibinfo {author} {\bibfnamefont {P.}~\bibnamefont
  {Klimas}},\ }\bibfield  {title} {\enquote {\bibinfo {title} {{Exact vortex
  solutions in a $CP^N$ Skyrme-Faddeev type model}},}\ }\href {\doibase
  10.1007/JHEP10(2010)008} {\bibfield  {journal} {\bibinfo  {journal} {JHEP}\
  }\textbf {\bibinfo {volume} {10}},\ \bibinfo {pages} {008} (\bibinfo {year}
  {2010})},\ \Eprint {http://arxiv.org/abs/1007.1667} {arXiv:1007.1667
  [hep-th]} \BibitemShut {NoStop}%
\bibitem [{\citenamefont {Ferreira}\ and\ \citenamefont
  {Leite}(1999)}]{Ferreira:1998zx}%
  \BibitemOpen
  \bibfield  {author} {\bibinfo {author} {\bibfnamefont {Luiz~A.}\ \bibnamefont
  {Ferreira}}\ and\ \bibinfo {author} {\bibfnamefont {Erica~E.}\ \bibnamefont
  {Leite}},\ }\bibfield  {title} {\enquote {\bibinfo {title} {{Integrable
  theories in any dimension and homogeneous spaces}},}\ }\href {\doibase
  10.1016/S0550-3213(99)00090-5} {\bibfield  {journal} {\bibinfo  {journal}
  {Nucl. Phys. B}\ }\textbf {\bibinfo {volume} {547}},\ \bibinfo {pages}
  {471--500} (\bibinfo {year} {1999})},\ \Eprint
  {http://arxiv.org/abs/hep-th/9810067} {arXiv:hep-th/9810067} \BibitemShut
  {NoStop}%
\bibitem [{\citenamefont {Piette}\ \emph
  {et~al.}(1995{\natexlab{a}})\citenamefont {Piette}, \citenamefont
  {Schroers},\ and\ \citenamefont {Zakrzewski}}]{Piette:1994ug}%
  \BibitemOpen
  \bibfield  {author} {\bibinfo {author} {\bibfnamefont {B.~M. A.~G.}\
  \bibnamefont {Piette}}, \bibinfo {author} {\bibfnamefont {B.~J.}\
  \bibnamefont {Schroers}}, \ and\ \bibinfo {author} {\bibfnamefont {W.~J.}\
  \bibnamefont {Zakrzewski}},\ }\bibfield  {title} {\enquote {\bibinfo {title}
  {{Multi - solitons in a two-dimensional Skyrme model}},}\ }\href {\doibase
  10.1007/BF01571317} {\bibfield  {journal} {\bibinfo  {journal} {Z. Phys. C}\
  }\textbf {\bibinfo {volume} {65}},\ \bibinfo {pages} {165--174} (\bibinfo
  {year} {1995}{\natexlab{a}})},\ \Eprint {http://arxiv.org/abs/hep-th/9406160}
  {arXiv:hep-th/9406160} \BibitemShut {NoStop}%
\bibitem [{\citenamefont {Piette}\ \emph
  {et~al.}(1995{\natexlab{b}})\citenamefont {Piette}, \citenamefont
  {Schroers},\ and\ \citenamefont {Zakrzewski}}]{Piette:1994mh}%
  \BibitemOpen
  \bibfield  {author} {\bibinfo {author} {\bibfnamefont {B.~M. A.~G.}\
  \bibnamefont {Piette}}, \bibinfo {author} {\bibfnamefont {B.~J.}\
  \bibnamefont {Schroers}}, \ and\ \bibinfo {author} {\bibfnamefont {W.~J.}\
  \bibnamefont {Zakrzewski}},\ }\bibfield  {title} {\enquote {\bibinfo {title}
  {{Dynamics of baby skyrmions}},}\ }\href {\doibase
  10.1016/0550-3213(95)00011-G} {\bibfield  {journal} {\bibinfo  {journal}
  {Nucl. Phys. B}\ }\textbf {\bibinfo {volume} {439}},\ \bibinfo {pages}
  {205--235} (\bibinfo {year} {1995}{\natexlab{b}})},\ \Eprint
  {http://arxiv.org/abs/hep-ph/9410256} {arXiv:hep-ph/9410256} \BibitemShut
  {NoStop}%
\bibitem [{\citenamefont {Hen}\ and\ \citenamefont
  {Karliner}(2008)}]{Hen:2007in}%
  \BibitemOpen
  \bibfield  {author} {\bibinfo {author} {\bibfnamefont {Itay}\ \bibnamefont
  {Hen}}\ and\ \bibinfo {author} {\bibfnamefont {Marek}\ \bibnamefont
  {Karliner}},\ }\bibfield  {title} {\enquote {\bibinfo {title} {{Rotational
  symmetry breaking in baby Skyrme models}},}\ }\href {\doibase
  10.1088/0951-7715/21/3/002} {\bibfield  {journal} {\bibinfo  {journal}
  {Nonlinearity}\ }\textbf {\bibinfo {volume} {21}},\ \bibinfo {pages}
  {399--408} (\bibinfo {year} {2008})},\ \Eprint
  {http://arxiv.org/abs/0710.3939} {arXiv:0710.3939 [hep-th]} \BibitemShut
  {NoStop}%
\bibitem [{\citenamefont {Karliner}\ and\ \citenamefont
  {Hen}(2010)}]{Karliner:2009at}%
  \BibitemOpen
  \bibfield  {author} {\bibinfo {author} {\bibfnamefont {Marek}\ \bibnamefont
  {Karliner}}\ and\ \bibinfo {author} {\bibfnamefont {Itay}\ \bibnamefont
  {Hen}},\ }\enquote {\bibinfo {title} {{Rotational Symmetry Breaking in Baby
  Skyrme Models}},}\ in\ \href {\doibase 10.1142/9789814280709_0008} {\emph
  {\bibinfo {booktitle} {{The multifaceted skyrmion}}}},\ \bibinfo {editor}
  {edited by\ \bibinfo {editor} {\bibfnamefont {Gerald~E.}\ \bibnamefont
  {Brown}}\ and\ \bibinfo {editor} {\bibfnamefont {Mannque}\ \bibnamefont
  {Rho}}}\ (\bibinfo  {publisher} {WORLD SCIENTIFIC},\ \bibinfo {year} {2010})\
  pp.\ \bibinfo {pages} {179--213},\ \Eprint {http://arxiv.org/abs/0901.1489}
  {arXiv:0901.1489 [hep-th]} \BibitemShut {NoStop}%
\bibitem [{\citenamefont {Adam}\ \emph {et~al.}(2009)\citenamefont {Adam},
  \citenamefont {Klimas}, \citenamefont {Sanchez-Guillen},\ and\ \citenamefont
  {Wereszczynski}}]{Adam:2009px}%
  \BibitemOpen
  \bibfield  {author} {\bibinfo {author} {\bibfnamefont {C.}~\bibnamefont
  {Adam}}, \bibinfo {author} {\bibfnamefont {P.}~\bibnamefont {Klimas}},
  \bibinfo {author} {\bibfnamefont {J.}~\bibnamefont {Sanchez-Guillen}}, \ and\
  \bibinfo {author} {\bibfnamefont {A.}~\bibnamefont {Wereszczynski}},\
  }\bibfield  {title} {\enquote {\bibinfo {title} {{Compact baby skyrmions}},}\
  }\href {\doibase 10.1103/PhysRevD.80.105013} {\bibfield  {journal} {\bibinfo
  {journal} {Phys. Rev. D}\ }\textbf {\bibinfo {volume} {80}},\ \bibinfo
  {pages} {105013} (\bibinfo {year} {2009})},\ \Eprint
  {http://arxiv.org/abs/0909.2505} {arXiv:0909.2505 [hep-th]} \BibitemShut
  {NoStop}%
\bibitem [{\citenamefont {Ferreira}\ \emph {et~al.}(2012)\citenamefont
  {Ferreira}, \citenamefont {Jaykka}, \citenamefont {Sawado},\ and\
  \citenamefont {Toda}}]{Ferreira:2011mz}%
  \BibitemOpen
  \bibfield  {author} {\bibinfo {author} {\bibfnamefont {L.~A.}\ \bibnamefont
  {Ferreira}}, \bibinfo {author} {\bibfnamefont {J.}~\bibnamefont {Jaykka}},
  \bibinfo {author} {\bibfnamefont {Nobuyuki}\ \bibnamefont {Sawado}}, \ and\
  \bibinfo {author} {\bibfnamefont {Kouichi}\ \bibnamefont {Toda}},\ }\bibfield
   {title} {\enquote {\bibinfo {title} {{Vortices in the Extended
  Skyrme-Faddeev Model}},}\ }\href {\doibase 10.1103/PhysRevD.85.105006}
  {\bibfield  {journal} {\bibinfo  {journal} {Phys. Rev. D}\ }\textbf {\bibinfo
  {volume} {85}},\ \bibinfo {pages} {105006} (\bibinfo {year} {2012})},\
  \Eprint {http://arxiv.org/abs/1112.1085} {arXiv:1112.1085 [hep-th]}
  \BibitemShut {NoStop}%
\bibitem [{\citenamefont {Klimas}\ and\ \citenamefont
  {Sawado}(2012)}]{Klimas:2012aw}%
  \BibitemOpen
  \bibfield  {author} {\bibinfo {author} {\bibfnamefont {Pawel}\ \bibnamefont
  {Klimas}}\ and\ \bibinfo {author} {\bibfnamefont {Nobuyuki}\ \bibnamefont
  {Sawado}},\ }\bibfield  {title} {\enquote {\bibinfo {title} {{Numerical
  vortex solutions in (3+1) dimensions for the extended $CP^N$ Skyrme-Faddeev
  model}},}\ }\href@noop {} {\  (\bibinfo {year} {2012})},\ \Eprint
  {http://arxiv.org/abs/1210.7523} {arXiv:1210.7523 [hep-th]} \BibitemShut
  {NoStop}%
\bibitem [{\citenamefont {Amari}\ \emph {et~al.}(2015)\citenamefont {Amari},
  \citenamefont {Klimas}, \citenamefont {Sawado},\ and\ \citenamefont
  {Tamaki}}]{Amari:2015sva}%
  \BibitemOpen
  \bibfield  {author} {\bibinfo {author} {\bibfnamefont {Yuki}\ \bibnamefont
  {Amari}}, \bibinfo {author} {\bibfnamefont {Pawel}\ \bibnamefont {Klimas}},
  \bibinfo {author} {\bibfnamefont {Nobuyuki}\ \bibnamefont {Sawado}}, \ and\
  \bibinfo {author} {\bibfnamefont {Yuta}\ \bibnamefont {Tamaki}},\ }\bibfield
  {title} {\enquote {\bibinfo {title} {{Potentials and the vortex solutions in
  the $CP^N$ Skyrme-Faddeev model}},}\ }\href {\doibase
  10.1103/PhysRevD.92.045007} {\bibfield  {journal} {\bibinfo  {journal} {Phys.
  Rev. D}\ }\textbf {\bibinfo {volume} {92}},\ \bibinfo {pages} {045007}
  (\bibinfo {year} {2015})},\ \Eprint {http://arxiv.org/abs/1504.02848}
  {arXiv:1504.02848 [hep-th]} \BibitemShut {NoStop}%
\bibitem [{\citenamefont {Arodz}\ \emph {et~al.}(2005)\citenamefont {Arodz},
  \citenamefont {Klimas},\ and\ \citenamefont {Tyranowski}}]{Arodz:2005gz}%
  \BibitemOpen
  \bibfield  {author} {\bibinfo {author} {\bibfnamefont {H.}~\bibnamefont
  {Arodz}}, \bibinfo {author} {\bibfnamefont {P.}~\bibnamefont {Klimas}}, \
  and\ \bibinfo {author} {\bibfnamefont {T.}~\bibnamefont {Tyranowski}},\
  }\bibfield  {title} {\enquote {\bibinfo {title} {{Field-theoretic models with
  V-shaped potentials}},}\ }\href@noop {} {\bibfield  {journal} {\bibinfo
  {journal} {Acta Phys. Polon. B}\ }\textbf {\bibinfo {volume} {36}},\ \bibinfo
  {pages} {3861--3876} (\bibinfo {year} {2005})},\ \Eprint
  {http://arxiv.org/abs/hep-th/0510204} {arXiv:hep-th/0510204} \BibitemShut
  {NoStop}%
\bibitem [{\citenamefont {Arodz}\ \emph {et~al.}(2007)\citenamefont {Arodz},
  \citenamefont {Klimas},\ and\ \citenamefont {Tyranowski}}]{Arodz:2007ek}%
  \BibitemOpen
  \bibfield  {author} {\bibinfo {author} {\bibfnamefont {H.}~\bibnamefont
  {Arodz}}, \bibinfo {author} {\bibfnamefont {P.}~\bibnamefont {Klimas}}, \
  and\ \bibinfo {author} {\bibfnamefont {T.}~\bibnamefont {Tyranowski}},\
  }\bibfield  {title} {\enquote {\bibinfo {title} {{Signum-Gordon wave equation
  and its self-similar solutions}},}\ }\href@noop {} {\bibfield  {journal}
  {\bibinfo  {journal} {Acta Phys. Polon. B}\ }\textbf {\bibinfo {volume}
  {38}},\ \bibinfo {pages} {3099--3118} (\bibinfo {year} {2007})},\ \Eprint
  {http://arxiv.org/abs/hep-th/0701148} {arXiv:hep-th/0701148} \BibitemShut
  {NoStop}%
\bibitem [{\citenamefont {Sawado}\ and\ \citenamefont
  {Yanai}(2021)}]{Sawado:2021rsc}%
  \BibitemOpen
  \bibfield  {author} {\bibinfo {author} {\bibfnamefont {Nobuyuki}\
  \bibnamefont {Sawado}}\ and\ \bibinfo {author} {\bibfnamefont {Shota}\
  \bibnamefont {Yanai}},\ }\bibfield  {title} {\enquote {\bibinfo {title}
  {{Phase analyses for compact, charged boson stars and shells harboring black
  holes in the $\mathbb{C}P^N$ nonlinear sigma model}},}\ }\href {\doibase
  10.1103/PhysRevD.103.125018} {\bibfield  {journal} {\bibinfo  {journal}
  {Phys. Rev. D}\ }\textbf {\bibinfo {volume} {103}},\ \bibinfo {pages}
  {125018} (\bibinfo {year} {2021})},\ \Eprint
  {http://arxiv.org/abs/2103.05877} {arXiv:2103.05877 [hep-th]} \BibitemShut
  {NoStop}%
\bibitem [{\citenamefont {Loginov}\ and\ \citenamefont
  {Gauzshtein}(2020)}]{Loginov:2020xoj}%
  \BibitemOpen
  \bibfield  {author} {\bibinfo {author} {\bibfnamefont {A.~Yu.}\ \bibnamefont
  {Loginov}}\ and\ \bibinfo {author} {\bibfnamefont {V.~V.}\ \bibnamefont
  {Gauzshtein}},\ }\bibfield  {title} {\enquote {\bibinfo {title} {Radially
  excited $u(1)$ gauged $q$-balls},}\ }\href {\doibase
  10.1103/PhysRevD.102.025010} {\bibfield  {journal} {\bibinfo  {journal}
  {Phys. Rev. D}\ }\textbf {\bibinfo {volume} {102}},\ \bibinfo {pages}
  {025010} (\bibinfo {year} {2020})}\BibitemShut {NoStop}%
\bibitem [{\citenamefont {Klimas}\ \emph {et~al.}(2022)\citenamefont {Klimas},
  \citenamefont {Sawado},\ and\ \citenamefont {Yanai}}]{Klimas:2022ghu}%
  \BibitemOpen
  \bibfield  {author} {\bibinfo {author} {\bibfnamefont {P.}~\bibnamefont
  {Klimas}}, \bibinfo {author} {\bibfnamefont {N.}~\bibnamefont {Sawado}}, \
  and\ \bibinfo {author} {\bibfnamefont {S.}~\bibnamefont {Yanai}},\ }\bibfield
   {title} {\enquote {\bibinfo {title} {Nodal compact $q$-ball and $q$-shell in
  the $\mathbb{C}{P}^{N}$ nonlinear sigma model},}\ }\href {\doibase
  10.1103/PhysRevD.105.085004} {\bibfield  {journal} {\bibinfo  {journal}
  {Phys. Rev. D}\ }\textbf {\bibinfo {volume} {105}},\ \bibinfo {pages}
  {085004} (\bibinfo {year} {2022})}\BibitemShut {NoStop}%
\end{thebibliography}%

\end{document}